\documentclass[lettersize,journal]{IEEEtran}

\usepackage{enumitem}  
\usepackage{cite}
\usepackage{amsmath,amssymb,amsfonts}
\usepackage{graphicx}
\usepackage{textcomp}
\usepackage{subfigure}
\usepackage{amsmath}

\newtheorem{definition}{Definition} 
\newtheorem{example}{Example} 
\usepackage{siunitx}
\usepackage{amsfonts}
\usepackage{comment}
\usepackage{amssymb}
\usepackage[table]{ xcolor}
\usepackage{multirow}
\usepackage{multicol}
\usepackage{array}
\usepackage{arydshln}
\usepackage{framed}
\usepackage{algpseudocode}
\usepackage{algorithm}
\def\BibTeX{{\rm B\kern-.05em{\sc i\kern-.025em b}\kern-.08em
    T\kern-.1667em\lower.7ex\hbox{E}\kern-.125emX}}
\begin{document}

\title{Kalman Filter Design for Intermittent Optical Wireless Communication Systems on Time Scales}
\author{Wenqi Cai, \IEEEmembership{Student Member, IEEE}, Bacem Ben Nasser, Mohamed Djemai, \IEEEmembership{Senior Member, IEEE}, \\and Taous Meriem Laleg-Kirati, \IEEEmembership{Senior Member, IEEE}
\thanks{Wenqi Cai, Bacem Ben Nasser, Taous Meriem Laleg-Kirati are with the Electrical and Computer Engineering Program, King Abdullah University of Science and Technology (KAUST), Thuwal 23955-6900, Saudi Arabia (e-mail: wenqi.cai@kaust.edu.sa; bacem.bennasser@kaust.edu.sa; taousmeriem.laleg@kaust.edu.sa).}
\thanks{Bacem Ben Nasser is with the Department of Mathematics, Physics and Computer Science, Higher Institute of Applied Sciences and Technology of Kairouan, University of Kairouan, Kairouan 3100, Tunisia (e-mail: bacem.bennasser@issatkr.rnu.tn).}
\thanks{Mohamed Djemai is with the Department of Automatic, University of Valenciennes and Hainaut-Cambresis, LAMIH, CNRS, UMR-8201, F-59313 Valenciennes, France (e-mail: mohamed.djemai@uphf.fr).}}



\maketitle

\begin{abstract}
Time-scale theory, due to its ability to unify the continuous and discrete cases, allows handling intractable non-uniform measurements, such as intermittent received signals. In this work, we address the state estimation problem of a vibration-induced intermittent optical wireless communication (OWC) system by designing a Kalman filter on time scales. First, the algorithm of the time-scale Kalman filter is introduced and a numerical example is given for illustration. Then the studied intermittent OWC system is presented, and experimental data are collected to determine the time scale's form, which has bounded graininess (a.k.a, bounded time jumps). Finally, we design a Kalman filter on the previously defined time scale for the intermittent OWC system and critically analyzed its estimation performance. Moreover, the obtained conclusions are further validated on a reference system. The simulation results corroborate that the time-scale Kalman filtering technique is considerably promising to solve the state estimation problem with non-uniform measurements. This study reveals for the first time the feasibility of applying the time-scale Kalman filter theory to practical applications.
\end{abstract}
\section{Introduction}
Optical wireless communication (OWC) is superior to traditional radio frequency (RF) communication in many aspects, such as it has super-high bandwidth, no license fee, no electromagnetic interference, etc., and is therefore widely studied and applied \cite{arnon2010underwater,cai2019modeling}. In OWC systems, one often adopts a narrow beam transmission configuration within the line-of-sight (LOS) setup. This setup not only saves energy but also ensures a more secure communication link. However, it is apparent that the LOS configuration depends on the precise alignment between the optical signal transmitter and the optical sensing receiver, which is even more challenging to achieve for slightly longer distance communication links \cite{cai2018robust}. Researchers have proposed some solutions to this problem, including $\mathcal {H} _ {\infty}$ pointing error control \cite{n2021reduction}, extremum seeking control \cite{cai2020extremum}, etc., in an attempt to solve the alignment problem of OWC systems. However, the control algorithms proposed in these works are only applicable to low-interference OWC systems and still need further improvement for the case of strong interference, where state estimation is a problem that needs to be addressed \cite{ solanki2018extended}. State estimation is a common and well-established technique to observe the internal state quantities of a system. However, implementing state estimation in alignment control of OWC systems is by no means an easy task. This is because, in practice, the optical communication links of OWC systems are often subject to a lot of interference, such as from waves, fish obstacles, and so on. Weaker interference can be eliminated by filtering techniques, but stronger interference can directly cause the receiver to lose signal completely from time to time, which is difficult to handle for state estimation. Such irregular intermittent reception signals are called \emph{non-uniform measurements} or \emph{intermittent measurements}, which are professionally defined as measurements that are not necessarily available at every constant time period \cite{isaza2018state}.

\par Recently, state estimation with intermittent measurements in networked systems has received increasing attention. Researchers have proposed a variety of observer structures and corresponding construction methods. In \cite{ferrante2016state}, the authors perform state estimation with respect to a linear time-invariant system with sporadic output measurements. A hybrid dynamics model interconnected with a jumping-fashion observer is considered, where the observer is activated by the arriving measurements. An improved approach is developed in \cite{sferlazza2018time} using the hybrid system formalism as well. Their strategy relies on solving a linear matrix inequalities problem with infinite dimensions. Similarly, the authors in \cite{berkane2019attitude} tackle the problem of designing attitude observers with measurement-triggered behavior in a hybrid framework. For nonlinear systems, two kinds of hybrid nonlinear observers with intermittent landmark position measurements for inertial navigation systems are introduced in \cite{wang2020nonlinear}. These two types rely on an infinite-dimensional optimization and a continuous-discrete Riccati equation, respectively. And in \cite{li2017robust}, the problem is studied in a distributed fashion networks.
\IEEEpubidadjcol
\par In Kalman filtering with intermittent measurements, researchers focus on analyzing the updated state covariance performance, as it defines the amount of uncertainty present in the system. Such works include \cite{sinopoli2004kalman}, which analyzes the statistical convergence characteristics in a discrete Kalman filter formulation for the estimation error covariance updating; \cite{ahmad2013extended}, which investigates the estimation uncertainty bound for a mobile robot system based on the extended Kalman filter; \cite{yang2017multi}, shows that the expected estimation error covariance for accessible sensors would diverge in a multi-sensor Kalman filtering setting as the transmission capacity is less than a threshold.

\par As presented, we broadly divide the above methods into two classes. One is the intuitive approach, that is to apply the hybrid formalism in which discrete and continuous time dynamics are coexisting. The other is to determine the bounds of state covariance to adjust the Kalman gain, such that the estimation performance is guaranteed. However, these methods are not straightforward because of the complexity of the design procedure, particularly the need for specific analyses of different systems, and the fact that they do not provide general answers for systems with non-uniform measurements of arbitrary bounded jumps (bounded graininess). Instead, an innovative idea to solve the state estimation problem with non-uniform measurements is to adopt the \emph{time scales theory}. Specifically, our interest is to fulfill the state estimation for an intermittent OWC system by designing a Kalman filter on time scales. 

\par
\begin{figure}[tbhp!]
\centering 
\includegraphics[width=0.4\textwidth]{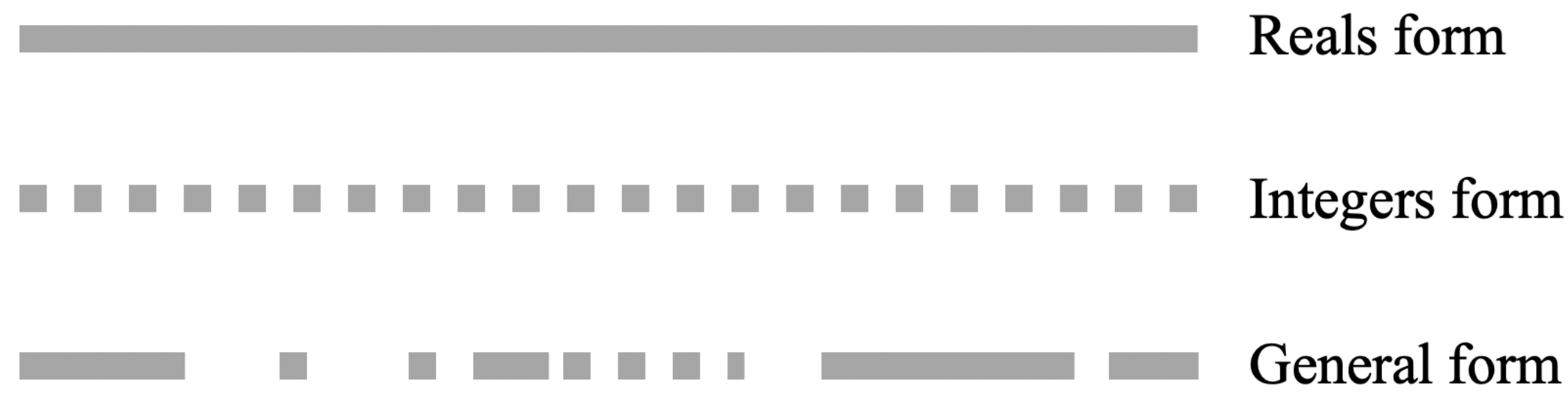}
\centering
\caption{Different forms of time scales}
\label{Tform}
\end{figure}

 In 1988, Dr. Stefan Hilger first formulated the time scales theory in his Ph.D. thesis \cite{hilger1988masskettenkalkul}. The main idea behind time scales theory is that it unifies discrete and continuous analysis. More specifically, it allows simultaneous operation of difference equations and differential equations, unifying these two kinds of equations to the so-called \emph{time scale dynamic equations}. As well known, many results within the framework of differential equations require complex derivations to be applied to difference equations, and in some cases, such extensions are even infeasible \cite{boyce2017elementary}. And time scales theory, studying the dynamic equations, is devoted to revealing and bridging such discrepancies. The field of time-scale dynamic equations encompasses, links, and extends the classical theory of difference and differential equations, hence the results obtained based on dynamic equations are of generalized characteristics. Therein, \emph{time scale}, an arbitrary nonempty closed subset in the field of real numbers, is the domain of the dynamic equations. Specifically, if the time scale is selected as a set of continuous real numbers, then the generalized solution of the dynamic equation is in fact a solution of an ordinary differential equation (ODE); similarly, the dynamic equation's generalized result is precisely the result of a difference equation if time scale is defined on integers. Generally, the typical form of the time scale is neither reals nor integers, but an arbitrary union of the both, as depicted in Fig.~\ref{Tform}. 

\par In addition to some theoretical studies of time scales theory in control \cite{nasser2018razumikhin,poulsen2019kalman,nasser2019sufficient,kumar2021finite}, this method has been employed to solve several engineering problems evolving especially on non-uniform time domains \cite{tisdell2008basic,chen2009periodic,zhou2011global,poulsen2019mean}, due to the unification and extension characteristics of time scales. These works illustrating the applicability of the theory in several fields such as economics, computer physics, population dynamics, etc. Nonetheless, most of the works on the time scales remain on the mathematical and theoretical sides. For instance, in the case of Kalman filtering on time scales, some progress has been made in its theory, while in terms of practical applications, no research has been conducted to verify the feasibility and validity of the theory.

\par In this paper, based on the theoretical algorithm of the time-scale Kalman filter proposed in \cite{bohner2013kalman}, we design a Kalman filter on time scales for an intermittent OWC system that suffers from packet loss issues. The idea is motivated by the following reasons:
\begin{itemize}
    \item The intermittent OWC system is a typical physical model characterized by continuous and discrete elements. And time scales theory, as a novel branch of mathematics, is undoubtedly an ideal candidate for achieving accurate modeling of intermittent OWC systems. This is because time scales theory allows a mathematical description of continuous-discrete hybrid processes using more general dynamical equations within a unified framework, rather than unilaterally using difference or differential equations.
    \item A conventional approach to solving the problem is the continuous-discrete Kalman filter, which consists of continuous-time state prediction and discrete-time state update \cite{kulikov2013accurate,mazenc2015continuous}. Yet, the limitation of this approach is that the discrete output must be uniformly sampled (i.e., with a fixed step size), which is often not the case for realistic intermittent OWC systems. In contrast, the time-scale Kalman filter perfectly suits the systems with non-uniform measurements of arbitrary bounded jumps (bounded graininess).
\end{itemize}
To the best of our knowledge, the time-scale Kalman filter has never been deployed to a real system to evaluate the estimation performance of the theory in practical engineering. The primary contributions of this paper are outlined below:
\begin{enumerate}
\item We provide a numerical analysis of the time-scale Kalman filter to facilitate the reader's understanding after giving the algorithm. As an extension of the numerical example in \cite{bohner2013kalman}, two plotting methods are presented to illustrate the estimation performance: plot in iteration and plot in time scale.
\item A linear model is derived for an intermittent OWC system subject to stochastic vibrations and therefore receives intermittent optical signals (non-uniform measurements).
\item Based on a specific form of time scale obtained from the experimental data, we design a Kalman filter on that time scale for the intermittent OWC system.
\item Simulation results indicate that Kalman filter on time scales is a decent candidate to resolve the state estimation problem for non-uniform measurements.
\end{enumerate}
Meanwhile, this study unveils the feasibility and current limitations of the time-scale Kalman filter theory in practical applications.

\par The manuscript is structured as follows. Some fundamental definitions about time scales theory are presented in Section \ref{sec2}. Section \ref{sec4} introduces the theory of the time-scale Kalman filter, including the algorithm and a numerical example. We then present the constructed intermittent OWC system in Section \ref{sec3}, which details the experimental setup, the system's model design, and the determination of the time scale form. Section \ref{sec5} elaborates the designing of the time-scale Kalman filter for the intermittent OWC systems. At last, conclusions are given in Section \ref{conclu}.
\section{Preliminaries}
\label{sec2}
In this section, to facilitate the reader's understanding of the subsequent sections, we briefly introduce a few of basic concepts about time scales theory. For more definitions and theorems, refer to \cite{bohner2001dynamic}.

\begin{definition}
A \emph{time scale} $\mathbb{T}$ is an arbitrary nonempty closed subset of the real numbers $\mathbb{R}$. Accordingly, $\mathbb{R}$, $\mathbb{Z}$, $\mathbb{N}$, and $\mathbb{N}_{0}$ are examples of time scales, while $\mathbb{Q}$, $\mathbb{R\setminus Q}$, $\mathbb{C}$, $(0,1)$ are not time scales.
\end{definition}
\begin{definition}
Let $\mathbb{T}$ be a time scale. For $t\in\mathbb{T}$ we define:
\begin{itemize}
\item the \emph{forward jump operator} $\sigma:\mathbb{T}$$\to$$\mathbb{T}$ by $\sigma{(t)}:=$inf\{$s\in\mathbb{T}:s>t$\};
\item the \emph{graininess function} $\mu:\mathbb{T}$$\to$$[0,\infty)$ by $\mu{(t)}:=\sigma{(t)}-t$;
\item let $f:\mathbb{T} \rightarrow \mathbb{R}$, the function $f^{\sigma}$ : $\mathbb{T}$$\to$$\mathbb{R}$ is given by $f^{\sigma}(t)=f(\sigma{(t)})$ for all $t\in\mathbb{T}$.
\end{itemize}
\end{definition}

\begin{definition}
\label{definition1}
    Assume a function $f:\mathbb{T}\to\mathbb{R}$, is continuous and $t\neq\sigma(t)$, then the \emph{delta derivative} of $f$ at $t$ is defined by $f^{\Delta}(t)$, with
\begin{equation*}
    f^{\Delta}(t)=\frac{f^\sigma(t)-f(t)}{\mu(t)}.
\end{equation*}
\end{definition}

\begin{example}
Let's consider the two special cases:
\begin{itemize}
\item[-] if $\mathbb{T}=\mathbb{R}$, $\sigma(t)=t$, $\mu(t)=\sigma(t)-t=0$,
\begin{equation*}
    f^{\Delta}(t)=\lim_{s \to t}\frac{f(t)-f(s)}{t-s}=f^{\prime}(t);
\end{equation*}
\item[-] if $\mathbb{T}=\mathbb{Z}$, $\sigma(t)=t+1$, $\mu(t)=\sigma(t)-t=1$,
\begin{equation*}
    f^{\Delta}(t)=\frac{f^\sigma(t)-f(t)}{\mu(t)}=\frac{f(t+1)-f(t)}{1}=\Delta f(t).
\end{equation*}
\end{itemize}
\end{example}
\section{Kalman Filter on Time Scales}
\label{sec4}
Since the 1970s, the Kalman filter has received great attention from both academic and industrial communities \cite{kim2018introduction}. As a mature technology, Kalman filter plays a key role in many engineering fields, e.g., state or parameter estimation, signal processing, and etc\cite{li2019novel}. In this work, the interest will be on the implementation of state estimation using Kalman filtering on arbitrary time scales.
\subsection{Algorithm Description}
\label{sec4.1}
The generalized Kalman filter on arbitrary time scales was introduced in \cite{bohner2013kalman} and the algorithm description is given in Table \ref{tab:timescale_algo}. From the algorithm, one could discover that the Kalman filter on time scales $\mathbb{T}$ has a similar structure with those in the classical time-domain (discrete and continuous domains) \cite{catlin2012estimation, anderson1971stability}, while two differences are emphasized here:
\begin{table*}[htbp!]
\renewcommand\arraystretch{1.3} 
    \centering
    \caption{The algorithm of time-scale Kalman filter \cite{bohner2013kalman}}
    \begin{tabular}{c|c}
    \hline
       System  &  $x^{\Delta}(t) = Ax(t) + Bu(t) + Gw(t)$, ${\rm{ x(}}{t_0}) = {x_0}$\\
       Measurement  &  $y(t) = Cx(t) + D + \upsilon (t)$\\
       \hdashline 
       Assumption  &  ${x_0} \sim \left( {{{\bar x}_0},{P_0}} \right)$, ${\rm{ }}\omega  \sim \left( {0,Q\delta (t, s)} \right)$, ${\rm{ }}\upsilon  \sim \left( {0,R\delta (t, s)} \right)$\\
       Initial estimate  &  $\hat x({t_0}) = {{\bar x}_0}$\\
       Error covariance  &  $P({t_0}) = {\rm E}\left[ {({x_0} - {{\hat x}_0}){{({x_0} - {{\hat x}_0})}^T}} \right] = {P_0}$ \\
    \hline  
    \hline 
     Estimate update &  ${\hat {x}^{\Delta}(t)} = A\hat x(t) + Bu(t) + K(t)[y(t) - c\hat x(t) - D]$\\
       Kalman gain  &  $K(t) = (I + \mu (t)A)P(t){C^T}{(R + \mu (t)CP(t){C^T})^{ - 1}}$\\
       Error covariance update  &  ${P^\Delta }(t) = AP(t) + (I + \mu (t)A)P(t){A^T} + GQ{G^T} - (I + \mu (t)A)P(t){C^T}{(R + \mu (t)CP(t){C^T})^{ - 1}}CP(t)(I + \mu (t){A^T})$\\
    \hline  
    \end{tabular}
    \label{tab:timescale_algo}
\end{table*}
\begin{figure}[tbp]
\centering
\subfigure[$t\in \mathbb{T}_1$: plot in iteration.]{
\begin{minipage}[t]{0.98\linewidth}
\centering
\includegraphics[width=3.25in]{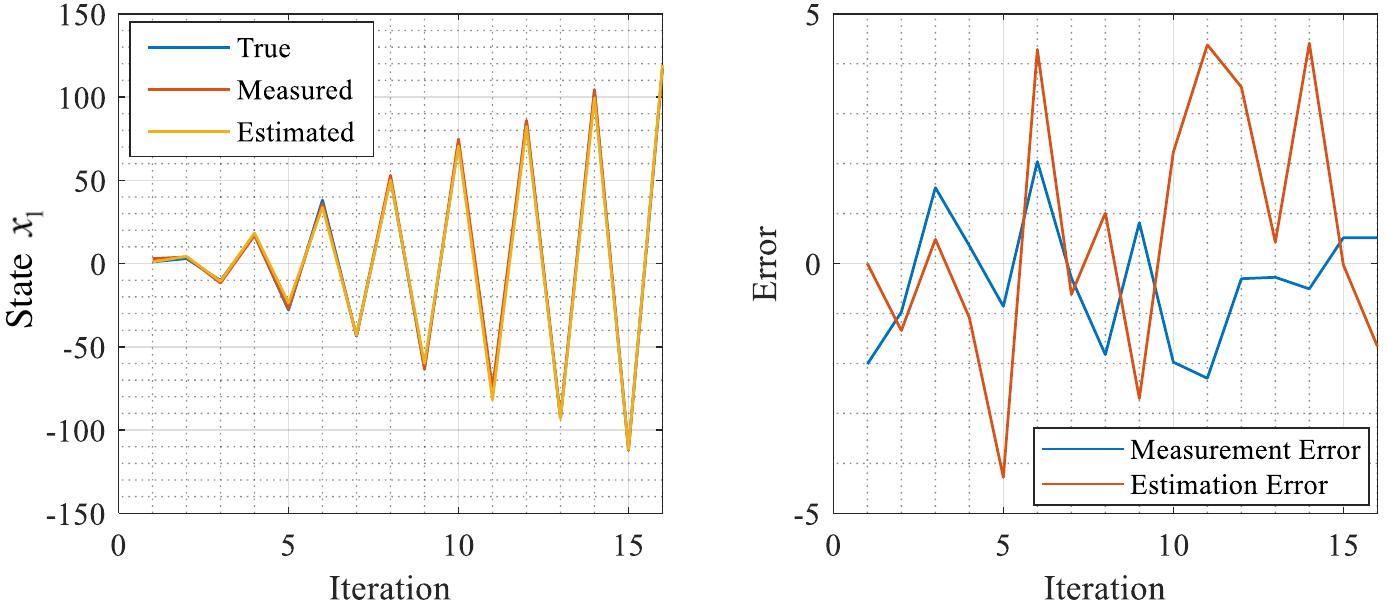}
\label{fig2.a}
\end{minipage}%
}%
\\
\subfigure[$t\in \mathbb{T}_1$: plot in time scale.]{
\begin{minipage}[t]{0.98\linewidth}
\centering
\includegraphics[width=3.25in]{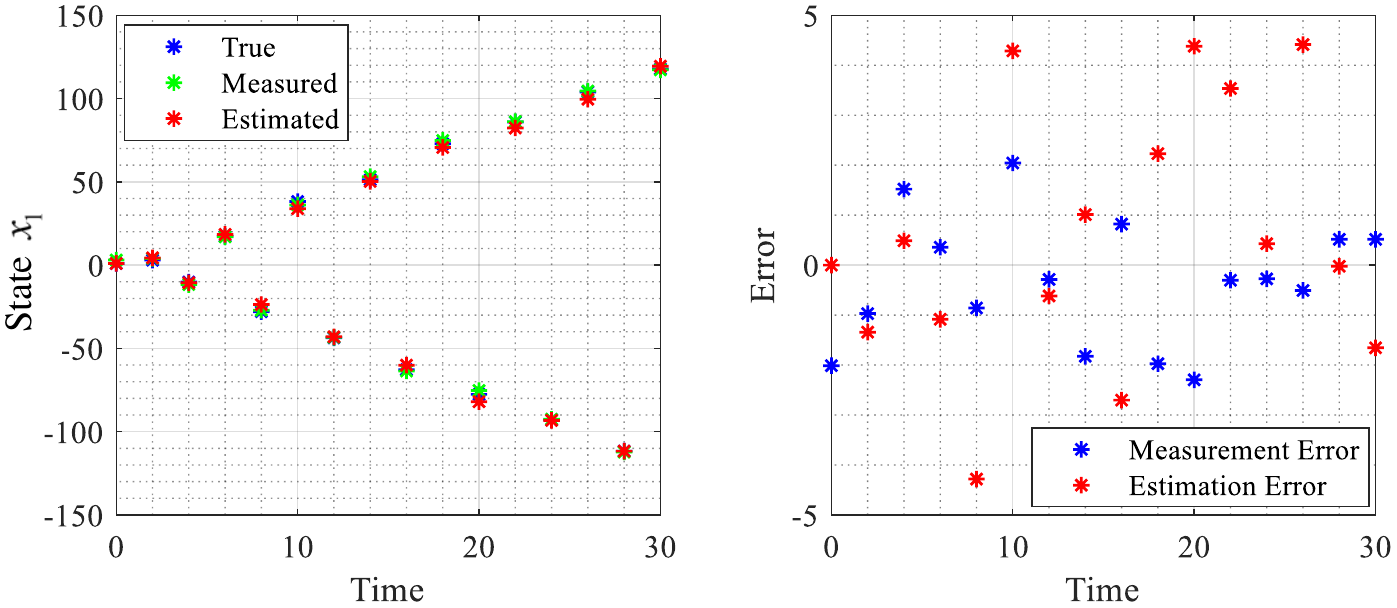}
\label{fig2.b}
\end{minipage}%
}%
\\
\subfigure[$t\in \mathbb{T}_2$: plot in iteration.]{
\begin{minipage}[t]{0.98\linewidth}
\centering
\includegraphics[width=3.25in]{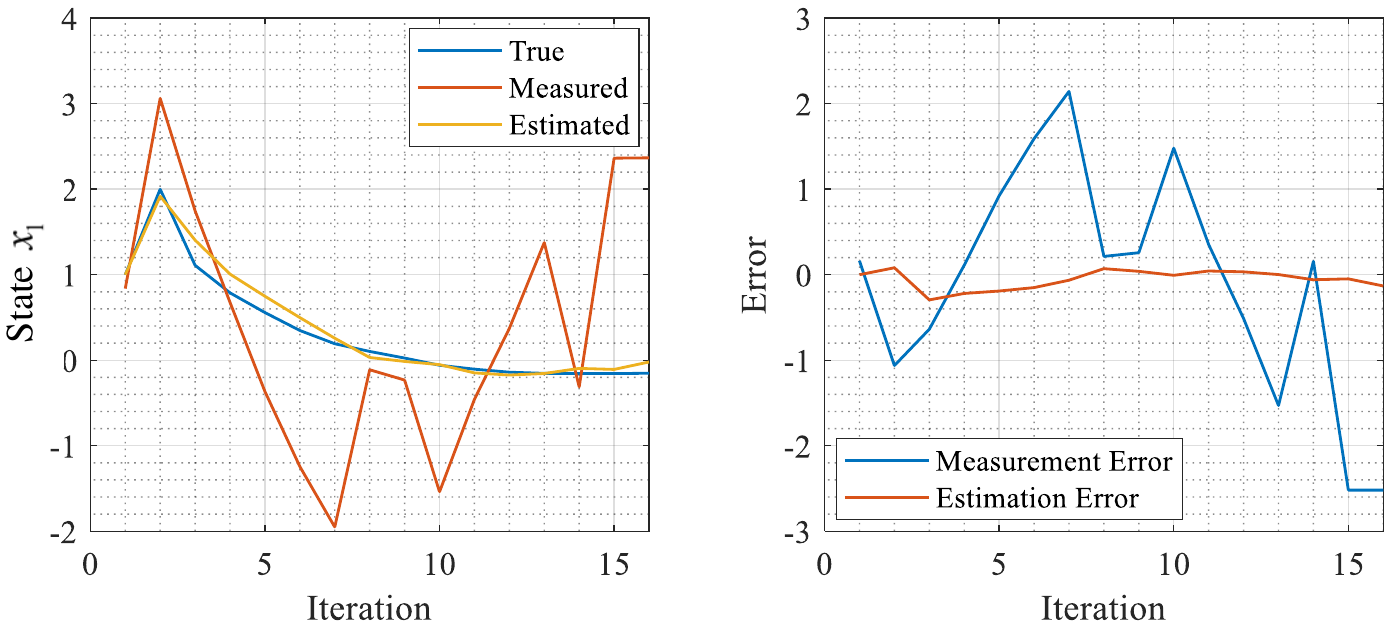}
\label{fig2.c}
\end{minipage}%
}%
\\
\subfigure[$t\in \mathbb{T}_2$: plot in time scale.]{
\begin{minipage}[t]{0.98\linewidth}
\centering
\includegraphics[width=3.25in]{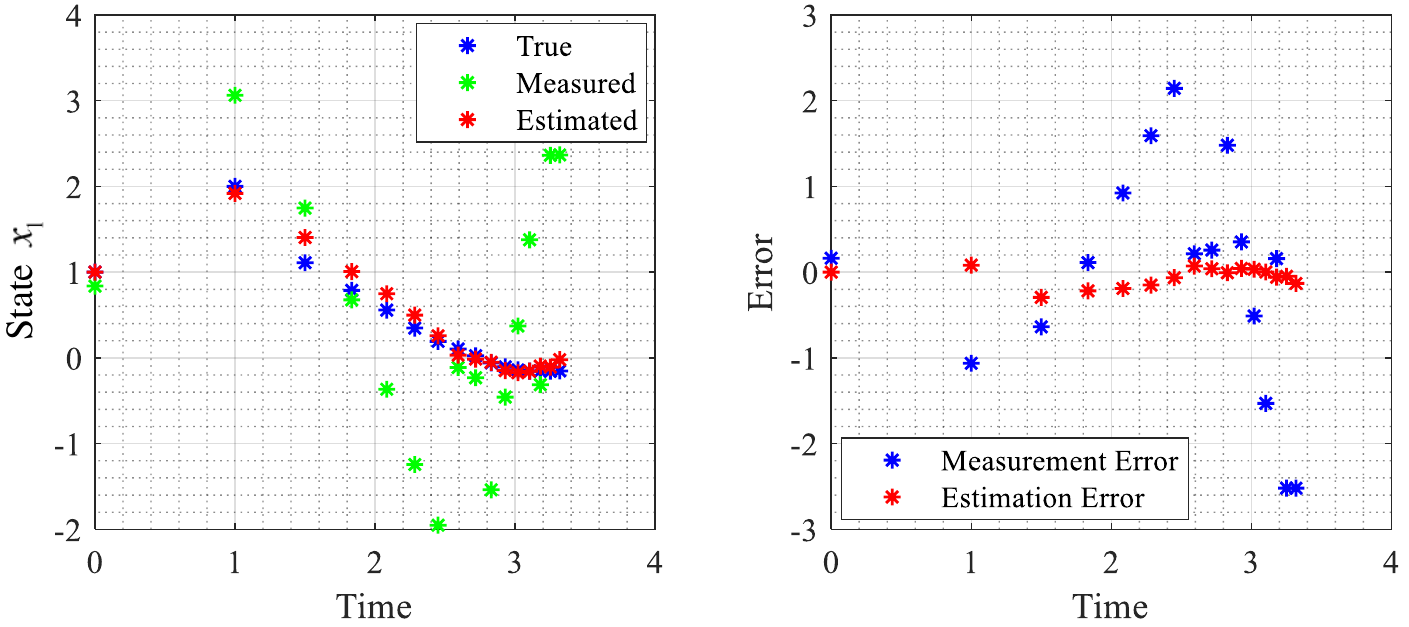}
\label{fig2.d}
\end{minipage}%
}%
\end{figure} 
\addtocounter{figure}{-1}    

\begin{figure}[tbp!]
\addtocounter{figure}{1} 
\subfigure[$t\in \mathbb{T}_3$: plot in iteration.]{
\begin{minipage}[t]{0.98\linewidth}
\centering
\includegraphics[width=3.25in]{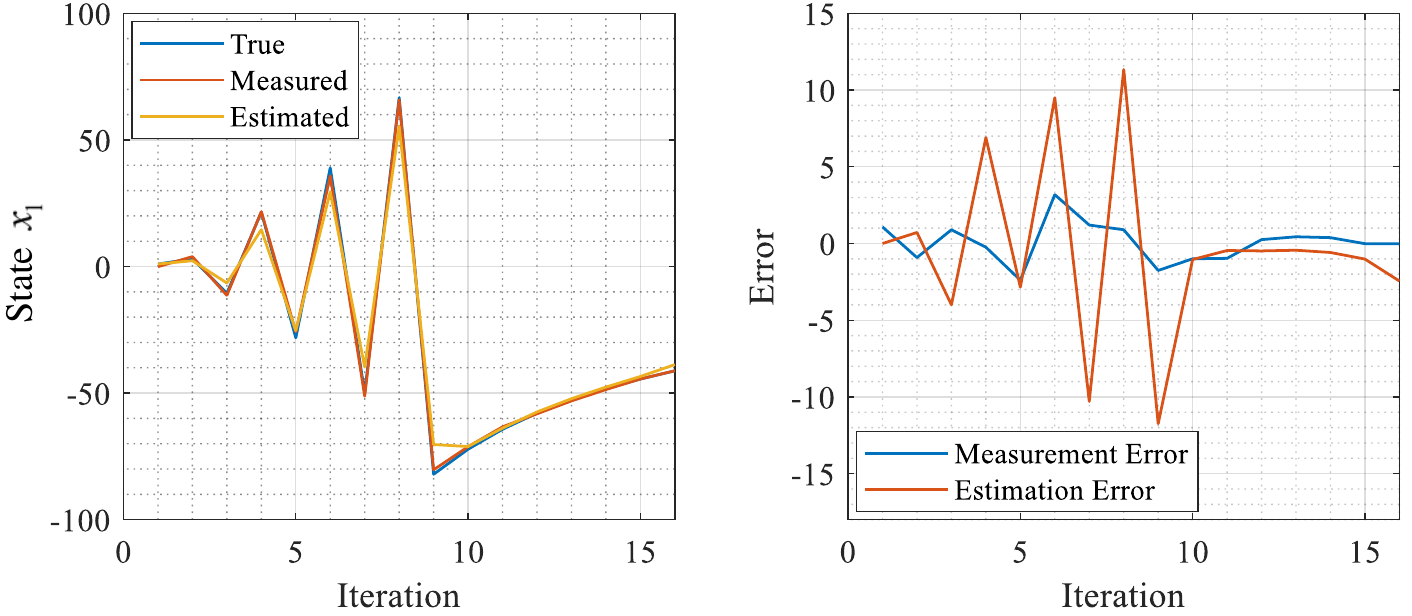}
\label{fig2.e}
\end{minipage}%
}%
\\
\subfigure[$ t\in \mathbb{T}_3$: plot in time scale.]{
\begin{minipage}[t]{0.98\linewidth}
\centering
\includegraphics[width=3.25in]{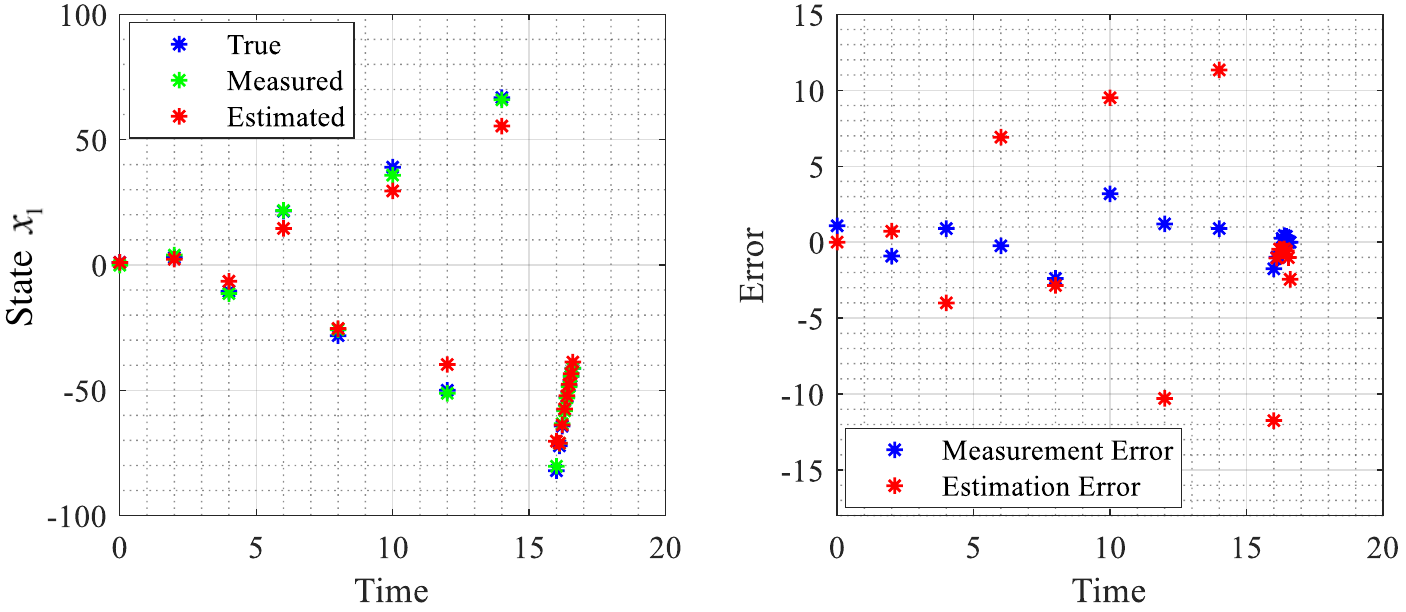}
\label{fig2.f}
\end{minipage}%
}%
\\
\subfigure[$t\in \mathbb{T}_4$: plot in iteration.]{
\begin{minipage}[t]{0.98\linewidth}
\centering
\includegraphics[width=3.25in]{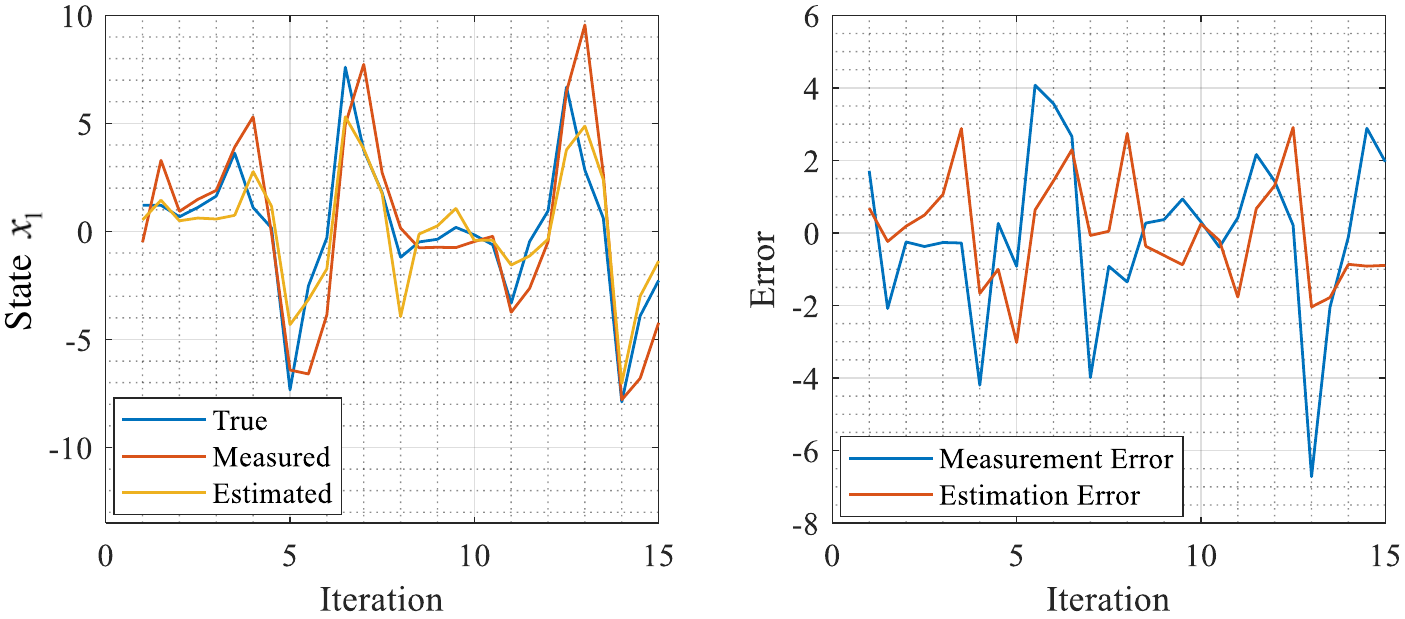}
\label{fig2.g}
\end{minipage}%
}%
\\
\subfigure[$t\in \mathbb{T}_4$: plot in time scale.]{
\begin{minipage}[t]{0.98\linewidth}
\centering
\includegraphics[width=3.25in]{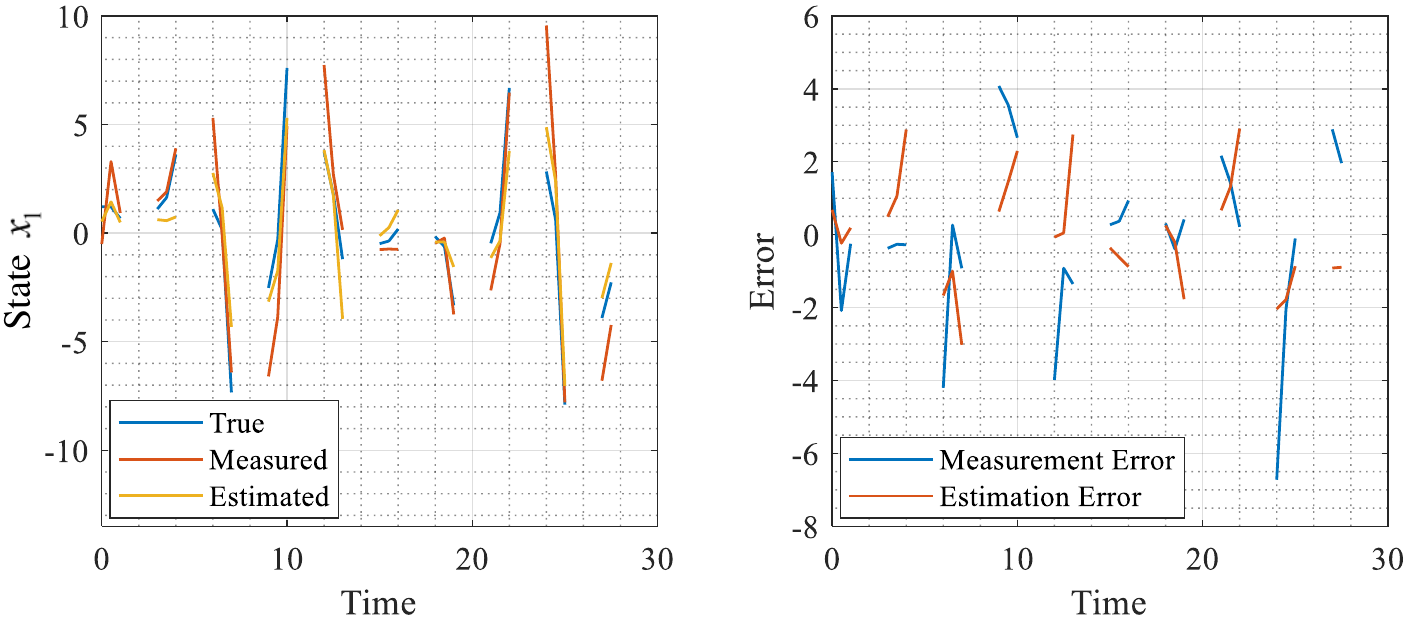}
\label{fig2.h}
\end{minipage}%
}%
\caption{State estimation using Kalman filter on different time scales.}
\label{fig:numerical example}
\end{figure}
\begin{itemize}
    \item The time-scale Kalman filter algorithm adopts dynamic equations ${x}^{\Delta}(t)$ instead of differential variations ${\dot x}(t)$ or difference equations $\Delta x(t)$ (for $P$ and $\hat {x}$ as well). Therefore, the generalized algorithm includes discrete and continuous Kalman filters as particular cases, and its structure or results can be extended to any non-uniform time domain.
    \item In the time-scale Kalman filter algorithm, the graininess function $\mu$ is involved in updating the Kalman gain and error covariance. Consequently, the Kalman filter on time scales depends heavily on $\mu$, that is, on the form of the time scales $\mathbb{T}$. (In fact, it also depends on the system, which will be discussed later in Section \ref{sec5}).
\end{itemize}

\subsection{Numerical Illustration}
\label{sec4.2}

In this part, the performance of the Kalman filter estimation on several different time sets is investigated for the following dynamics \cite{bohner2013kalman}:
\begin{equation}
\label{equ1}
\begin{cases}
  {x^\Delta }(t) = Ax(t) + Bu(t) + G\omega(t)  & \text{ } \\
  y = Cx(t) + D + \upsilon(t)   & 
  \end{cases},
\end{equation}
where $x=(x_1,x_2)^{\top}\in \mathbb{R}^{2}$ is the state vector and $u=(u_1,u_2)^{\top}\in \mathbb{R}^2$ is the input vector. Variable $w \sim N\left( {0,1} \right)$ is the normally distributed process noise and $v  \sim N\left( {0,2} \right)$ denotes the normally distributed measurement noise. Besides, we have
\begin{equation*}
A = \left[ {\begin{array}{*{20}{c}}
0&1\\
{ - 1}&{ - 2}
\end{array}} \right],\;B=D=0,\;G = \left[ {\begin{array}{*{20}{c}}
0\\
1
\end{array}} \right],\;C = \left[ {\begin{array}{*{20}{c}}
1\\
0
\end{array}} \right]^{\top}.
\end{equation*}
The values of the state, state estimation, and error covariance are initialized by
\begin{equation*}
{\bar x}_0 = {\hat x}_0 = \left[ {\begin{array}{*{20}{c}}
1\\
1
\end{array}} \right]\hbox{ and }P_0 = \left[ \begin{array}{*{20}{c}}
2&0\\
0&3
\end{array}\right].
\end{equation*}
Applying the considered algorithm, we design Kalman filters for system (\ref{equ1}) on four time scales that have bounded graininess functions $\mu(t)$: $\mathbb{T}_1=2\mathbb{Z}$, $\mathbb{T}_2=(H_n)_{n \in \mathbb{N}_0}$, $\mathbb{T}_3=\begin{cases}
     2\mathbb{Z} & \hbox{ if }t\leq8 \\
   \mathbb{T}_2   & \hbox{if } t>8
\end{cases}$ and
 $\mathbb{T}_4=P_{1,2}$, where $ H_n= \sum_{k=1}^{n}\frac{1}{k}$ with $n\in\mathbb{N}_0$ and $ P_{1,2}= \bigcup\limits_{k= 0}^\infty {[3k,3k+1]} $. It can be derived that for $t \in \mathbb{T}_1$, one get $\mu(t)=2$, for $t \in \mathbb{T}_2$, one get $\mu(t)=\frac{1}{n+1}$, for $t \in \mathbb{T}_3$, one get $\mu(t)=\begin{cases}
    2  & \hbox{ if }t\leq8  \\
   \frac{1}{n+1}   & \hbox{if } t>8
\end{cases}$ and for $t\in \mathbb{T}_4$, one get $ \mu(t)=\begin{cases}
    0  & \hbox{ if } t \in \bigcup\limits_{k = 0}^\infty  {\left[ {3k,{\rm{ }}3k+1} \right)}  \\
   2   & \text{otherwise}
\end{cases}$.\\

\par In system (\ref{equ1}), the output is $y=x_1+\upsilon$. The state $x_1$ is measurable and is therefore chosen as a reference for the estimation performance of the designed Kalman filter. Figure \ref{fig:numerical example} contrastingly shows the numerical simulations of the time-scale Kalman filter of (\ref{equ1}) on time sets $\mathbb{T}_{i}$ ( $i\in\{1,\dots,4\}$) using two plotting methods. The first method, ``\emph{Plot in iteration}", refers to using the number of iterations as the abscissa when plotting the results. This means that the Kalman filter operates on the time scale as an iterative calculation according to the algorithm, where only the value of the graininess $\mu$ at time $t$ is considered, regardless of how the real time is represented. As a result, the plotted results are continuous graph lines, from which no information of the time scale $\mathbb{T}$ can be obtained (see Fig. \ref{fig2.a}, Fig. \ref{fig2.c}, Fig. \ref{fig2.e} and Fig. \ref{fig2.g}). The second one, ``\emph{Plot in time scale}", uses real-time as the abscissa, so the form of time scale can be directly seen from the resulting graph (see Fig. \ref{fig2.b}, Fig. \ref{fig2.d}, Fig. \ref{fig2.f} and Fig. \ref{fig2.h}). It is worth being aware that although the graphical results of ``plot in time scale" look more intuitive, the essence of the calculation remains the iterative method.

\par From Fig.~\ref{fig:numerical example}, one can observe that:
\begin{itemize}
    \item When $t\in \mathbb{T}_1$, the Kalman filter has a good estimation, as can be seen from the left-hand plots of Fig.~\ref{fig2.a}, Fig.~\ref{fig2.b}. However, the estimation error is a bit larger than the measurement error, which is not usually the case in the classical time domain (discrete and continuous domains).
    \item When $t \in\mathbb{T}_2$, the estimation of the Kalman filter is sufficiently accurate. Moreover, the estimation error is much smaller than the measurement error.
    \item When $t\in \mathbb{T}_3$, the conclusion is analogous to the first case.
    \item When $t\in\mathbb{T}_{4}$, the performance of the Kalman filter is slightly worse, and the estimation error is somewhat smaller than the measurement error.
\end{itemize}
In conclusion, the estimation performance of the time-scale Kalman filter is closely related to the time scale $\mathbb{T}$. Although the estimation effect varies with the form of $\mathbb{T}$, it is satisfactory in most cases. Moreover, the absolute value of the estimation error seems to be bounded in all cases, which is crucial when considering practical applications. Furthermore, it should be mentioned that although the values of the measurement error and estimation error depend heavily on the values of the measurement noise $\upsilon$ and the system process noise $\omega$, the estimation effect of system (\ref{equ1}) at a particular $\mathbb{T}$ is consistent. For instance, the estimation error is always greater than the measurement error when $\mathbb{T}=\mathbb{T}_{1}$ for all appropriate noise levels.
\section{Intermittent OWC System}
\label{sec3}
This section describes the constructed intermittent OWC system, including the experimental setup, the model design of the system, and the determination of the time scale form.
\subsection{Experimental Setup}
\label{sec3.1}
\begin{figure}[tbp]
\centering 
\includegraphics[height=1.8in,width=3.2in]{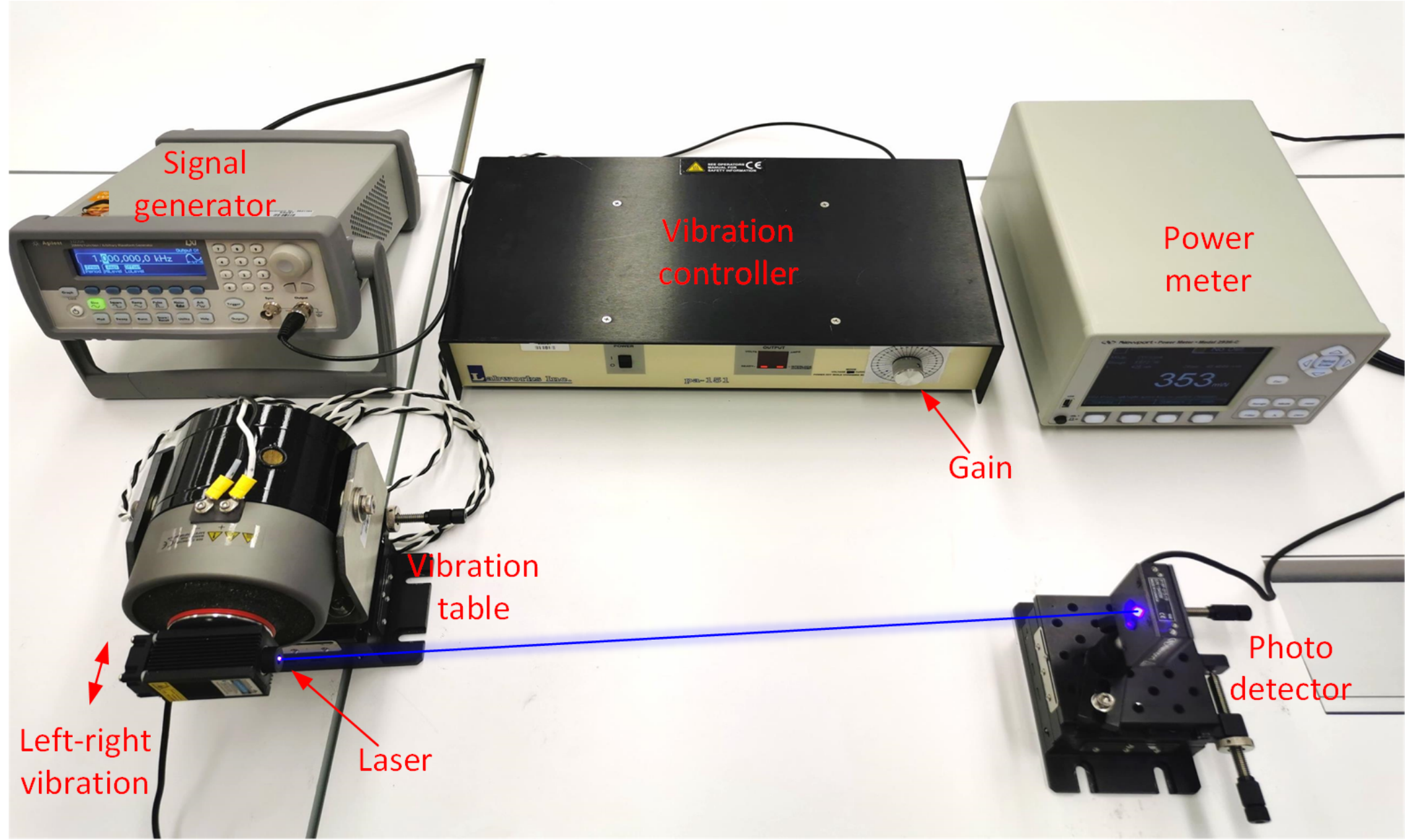}
\centering
\caption{Experimental setup of the vibration-induced intermittent OWC system.}
\label{fig:setup_4}
\end{figure}

The experimental setup of the vibrating optical communication system is shown in Fig.~\ref{fig:setup_4}. An ordinary collimating laser is used as the transmitter, which is mounted on a vibration table, performing horizontally left-right vibrations along with the vibration table. The vibration signal comes from a signal generator (we adopt, in this work, a random signal with a frequency of $2$\si{Hz}). The generated signal is controlled by the ``gain" knob of the vibration controller to regulate the amplitude (intensity) of the vibration. The photodetector (PD) acts as a receiver, and the power of the received optical signals is continuously recorded by a power meter. When the vibration amplitude is high enough, the received optical power decays to zero. Therefore, with a random vibration signal, an intermittent optical signal (non-uniform measurement) is acquired. 
\subsection{Model Design}
\label{sec3.2}
\begin{figure}[tbp]
\centering 
\includegraphics[width=0.90\linewidth]{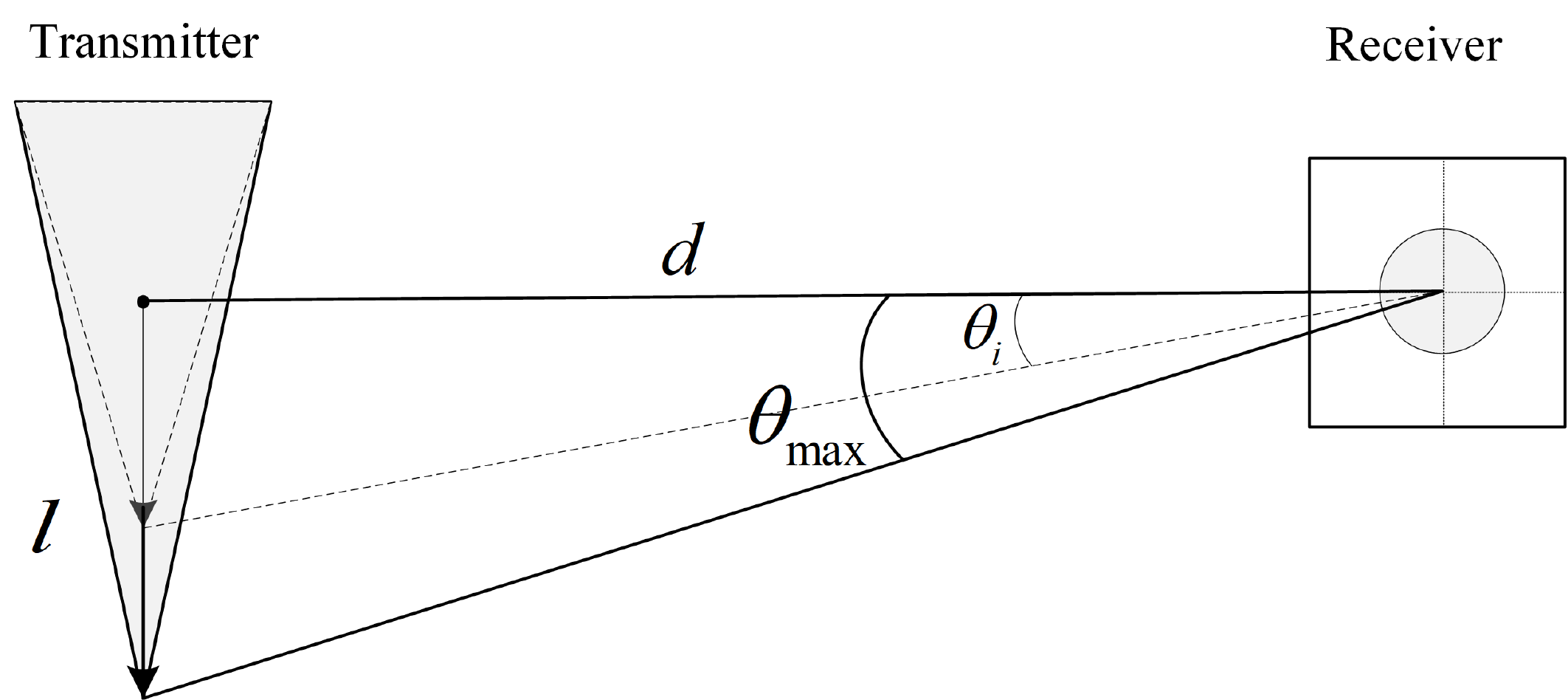}
\centering
\caption{Simplified model of the relative motion between the laser and the PD.}
\label{fig:config_t}
\end{figure}

Based on the experimental setup diagram, a simplified model is obtained as shown in Fig.~\ref{fig:config_t}, which illustrates the relative motion between the transmitter (laser) and the receiver (PD). The laser vibrates from the origin to the right with a maximum vibration amplitude of $l$, which is corresponded to the maximum relative angle $\theta_{\max}$ between the receiver and the transmitter (the link spacing is $d$). When the laser is at a certain position within $l$, the relative angle is $\theta_{i}$ and the received optical power ($P_i$) at each angle $\theta_{i}$ is measured. Obviously, the maximum power ($P_{\max}=270$\si{mW}) occurs when the laser is at the origin, and the minimum power ($P_{min}=0$\si{mW}) occurs when the laser vibrates to the far right.
\begin{figure}[t]
\centering
\subfigure[Received power vs. Relative angle]{
\begin{minipage}[t]{0.9\linewidth}
\centering
\includegraphics[width=2.8in]{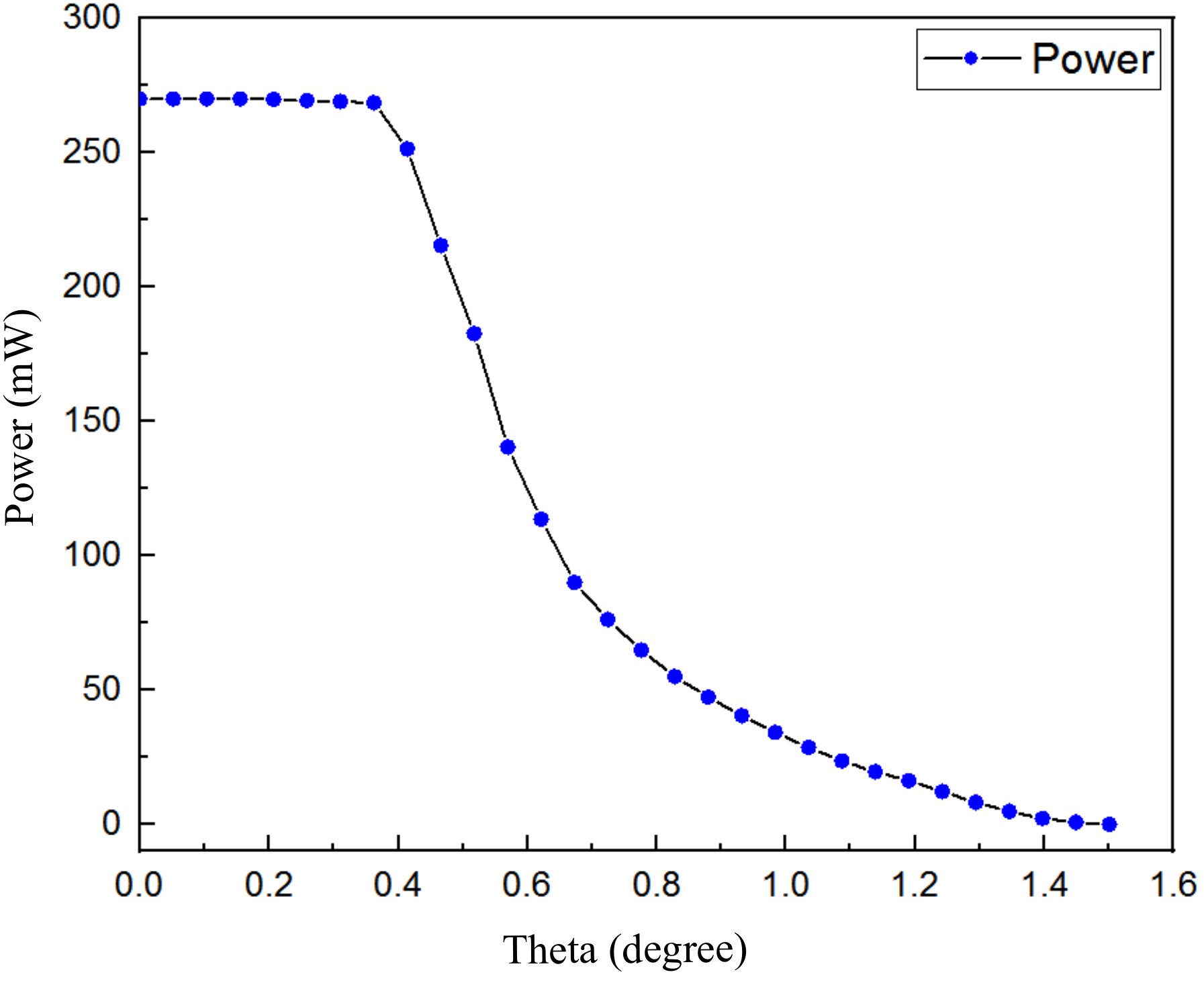}
\label{allpart}
\end{minipage}%
}%
\\
\subfigure[Extract the linear part of (a)]{
\begin{minipage}[t]{0.9\linewidth}
\centering
\includegraphics[width=2.8in]{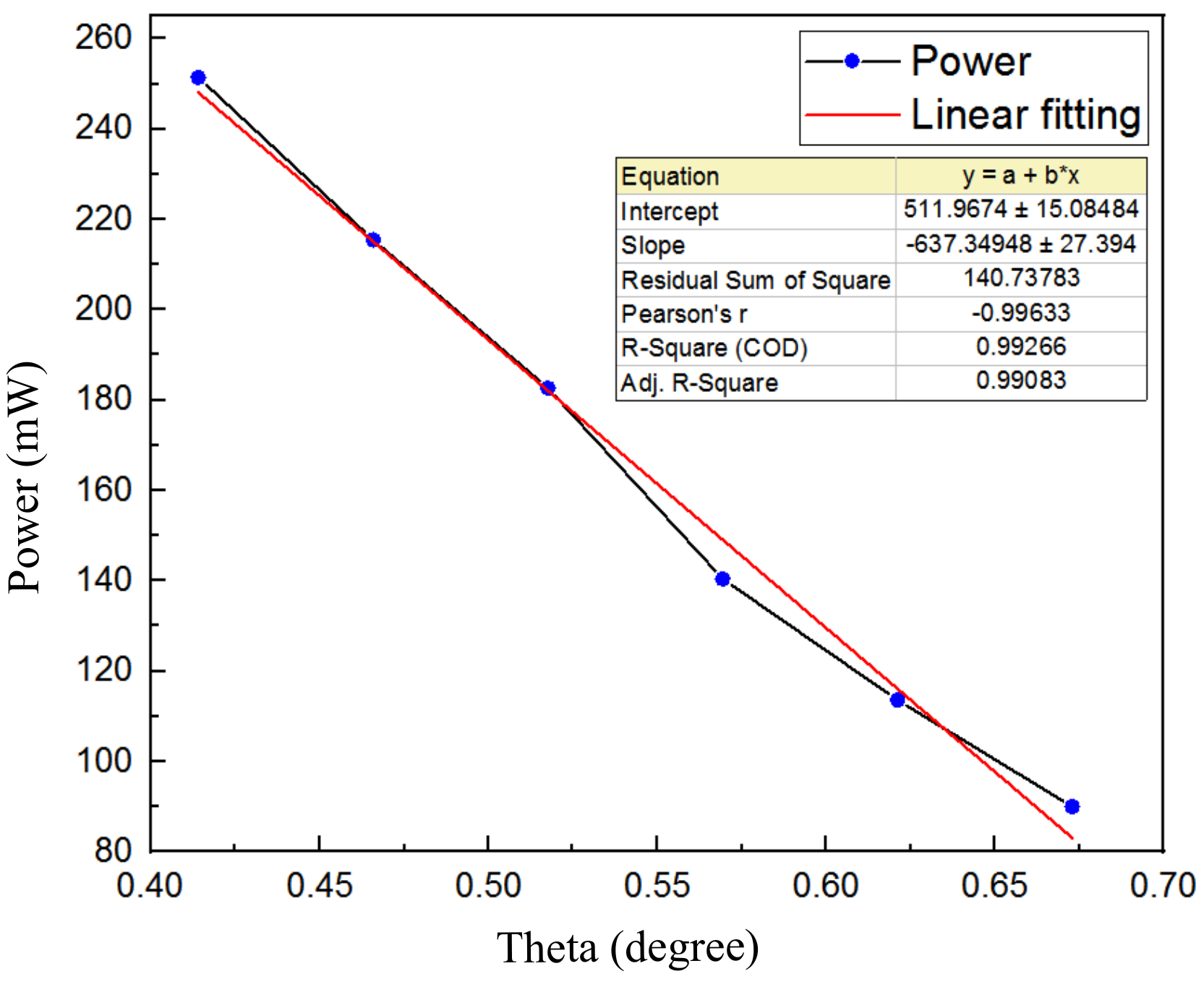}
\label{linearpart}
\end{minipage}%
}%
\centering
\caption{Relation mapping between the received power and the relative angle.}
\label{fig:part}
\end{figure}

\par The relation mapping with $30$ sampling points between the received optical power $P_i$ and the relative angle $\theta_i$ is shown in Fig.~\ref{fig:part}. One can observe that the optical power drops quite smoothly when the angle is in the range of $[0,0.4]\cup[0.7,1.51]$. This can be explained by the fact that the photodetector has a specified detection area, and the power of the moving optical spot within the detection area is almost invariant. When the optical spot is completely moved out of the detection area, the optical power changes slowly likewise due to the low-power radiance of the optical spot. However, while moving the optical spot out of the detection area, the received power will decay rapidly and approximately linearly. The linear portion intercepted from Fig.~\ref{allpart} is shown enlarged in Fig.~\ref{linearpart}, with $\theta_i$ ranging in the interval $[0.4141,0.6729]$. Since the slope of this linear part is the largest, the optical power varies most dramatically with the relative angle. Therefore, we are more interested in designing the Kalman filter within this range.

\par From the linear fitting equation shown in Fig.~\ref{linearpart}, the relation mapping formula between $P_i$ and $\theta_i$ is written as
\begin{equation}
    P_i(\theta_i) = -637.35  \theta_i + 511.97.
\end{equation}
We choose the system state and output as
\begin{equation*}
   \left\{ {\begin{array}{*{20}{c}}
{ x = \theta_i}\\
\begin{array}{l}
y = P_i,
\end{array}
\end{array}} \right.
\end{equation*}
as a result, the dynamic equation of the intermittent OWC system on time scales can be expressed as
\begin{equation}
\left\{ {\begin{array}{*{20}{l}}
{x^{\Delta}(t)=x(t)+\omega(t)}\\
{y(t) = -637.35 x(t) + 511.97 + \upsilon(t)}
\end{array}} \right..
\label{systemt}
\end{equation}
Variable $\omega$ refers to the process noise, which in this case represents the increments or decrements of the relative angle induced by the random (Gaussian) vibrations. From the system's operational perspective, $\omega$ is the amount of randomly turning the ``gain" knob of the vibration controller, obeying $\omega\sim N(0,1)$. And $\upsilon$ is the measurement noise with $\upsilon\sim N(0,100)$.
\subsection{Determine the Time Scale}
\label{sec3.3}
\begin{figure}[tbp]
\centering
\subfigure[``Time-power" experimental graph]{
\begin{minipage}[t]{0.98\linewidth}
\centering
\label{powert experiment graph}
\includegraphics[width=3.3in]{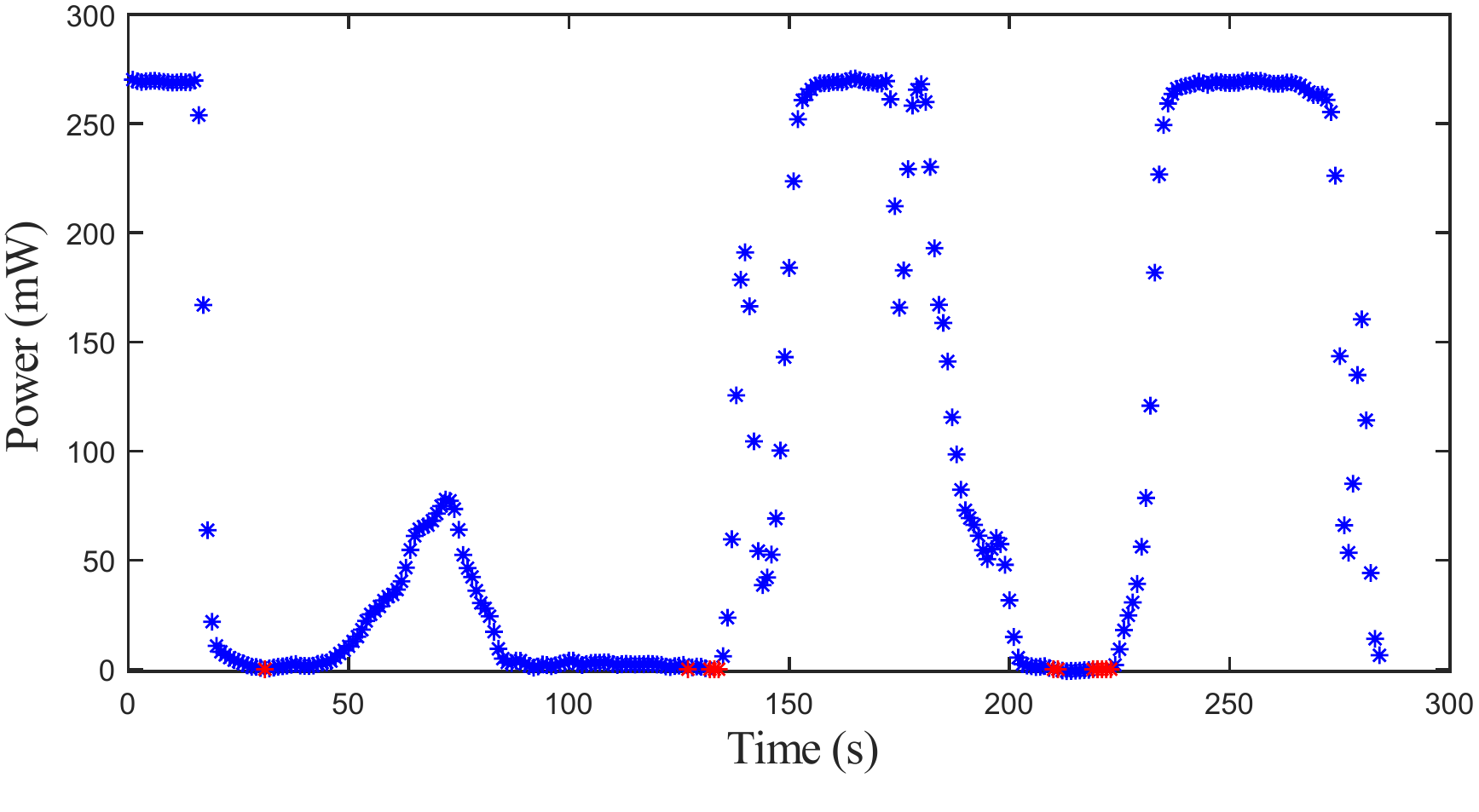}
\end{minipage}%
}%
\\
\subfigure[Redraw graph (a)]{
\begin{minipage}[t]{0.98\linewidth}
\centering
\label{Redraw grapha}
\includegraphics[width=3.2in]{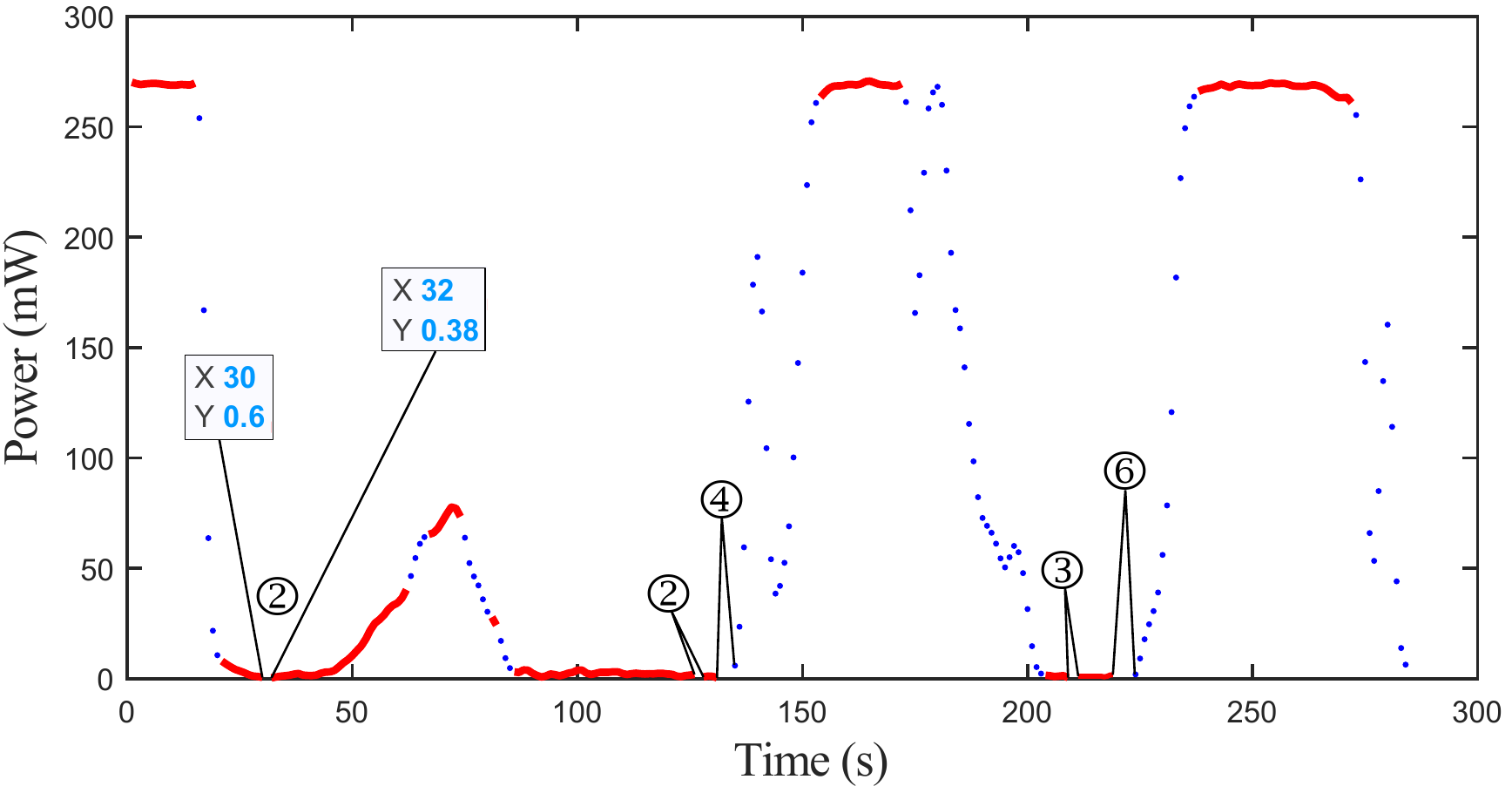}
\end{minipage}%
}%
\caption{Time set extraction.}
\label{fig:power_t}
\end{figure}
After deriving the time-scale model of the system, the objective of this subsection is to determine a specific time scale $\mathbb{T}_d$ based on the experimental data. As mentioned before, the ``gain" knob is randomly rotated to vary the amplitude of vibration while recording the received optical power (once per second). Figure \ref{powert experiment graph} shows a $5$-minute ($300$ seconds) ``time-power" experimental graph, where blue dots represent the valid data and red dots represent the invalid data. ``Invalid" means that the received optical power is equal to or less than zero, i.e., the PD is not receiving any signal at this moment, which is regarded as a signal interruption (the corresponding signal is called intermittent signal). In the rest of the paper, we name the length of the discontinued interval of the intermittent signal as a time jump $J$. In addition, some blue data points are densely distributed in the figure, while others are sparsely distributed. Therefore, by considering the dense part as continuous, the sparse part as discrete, and the invalid part as time jump, the ``time-power" plot can be redrawn as shown in Fig.~\ref{Redraw grapha}.

\par The numbers in the circles in Fig.~\ref{Redraw grapha} indicate the values of the five time jumps $J$. Taking ``\textcircled{2}" as an example, when $t = 30\si{s}$, the subsequent signal is interrupted for two seconds, i.e, $\mu(30)=2$, so the time sequence is ``$ \ldots, 30] \cup [32, \ldots$". As seen in the figure, the longest time jump is $J = 6$, which means that the graininess $\mu(t)$ of the time scale is bounded with $0 \leq \mu(t) \leq 6$. Besides, the red lines signify the continuous time and the blue dots signify the discrete time. In summary, the time scale $\mathbb{T}_d$ with bounded graininess is determined as
\begin{equation*}
\begin{aligned}
    \mathbb{T}_{d}=&[1,15)\cup\{15, \ldots, 21\} \cup (21, 30] \cup [32,62)\cup\{62, \ldots, \\
    &67\}\cup(67,75)\cup\{75, \ldots, 81\}\cup(81,83)\cup\{83, \ldots, 86\}\\
    &\cup(86, 126] \cup [128,131]\cup\{135, \ldots, 154\}\cup(154,173)\\
    &\cup\{173, \ldots, 204\}\cup(204, 209] \cup [212,218]\cup\{224, \ldots,\\
    &238\}\cup (238,273)\cup\{273, \ldots, 285\}\cup (285,300].
\end{aligned}
\end{equation*}
\section{Design Kalman Filter on Time Scales}
\label{sec5}
\begin{figure*}[hbtp]
\centering 
\includegraphics[width=0.88\textwidth]{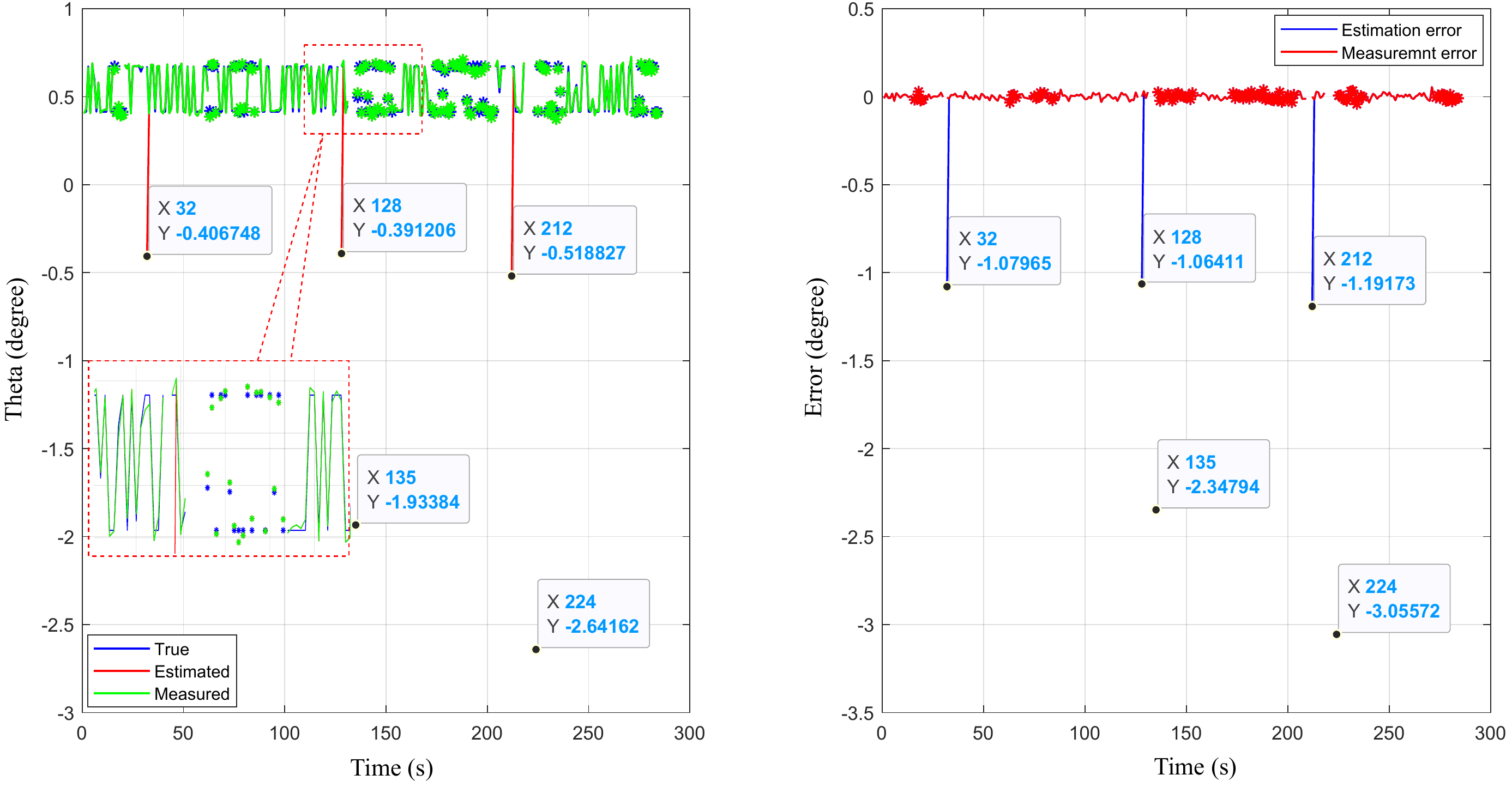}
\centering
\caption{The result of Kalman filter for the intermittent OWC system (\ref{systemt}) on $\mathbb{T}_d$.}
\label{fig:kf1}
\end{figure*}
\begin{figure*}[hbtp]
\centering 
\includegraphics[width=0.88\textwidth]{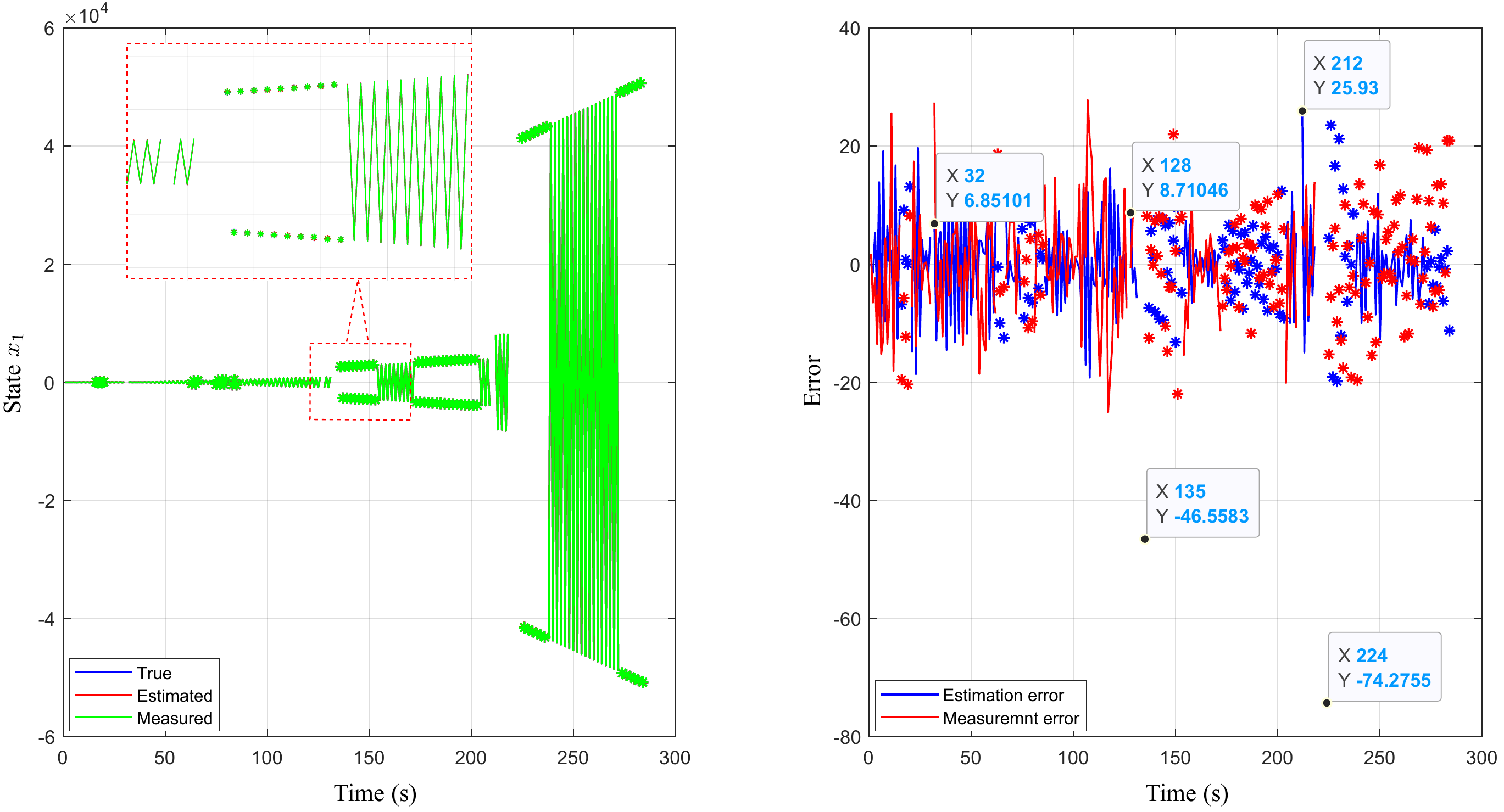}
\centering
\caption{The result of Kalman filter for the reference system (\ref{equ1}) on $\mathbb{T}_d$.}
\label{fig:kf2}
\end{figure*}

\par In this section, the implementation of Kalman filter for the intermittent OWC system on time scales is elaborately examined, exposing the potential capabilities and restrictions of applying time scales theory. The design of Kalman filter for the derived OWC system (\ref{systemt}) on $\mathbb{T}_d$ follows the algorithm in Table \ref{tab:timescale_algo} as well, whose result is shown in Fig.~\ref{fig:kf1}.

\par The results are plotted in the time scale form directly. As can be observed, the estimation errors are much larger than the measurement errors when $t \in\{32,128,135,212,224\}$. Additionally, except for the five marked points, both the measurement and the estimation are very close to the true state $\theta$ (the three lines almost overlap). In other words, the Kalman filter seems to perform well if the five points (which we call problematic points) are not taken into account. Interestingly, from the time scale form $\mathbb{T}_d$, it can be found that these five points are exactly the time points after each time jump $J$. This is essentially due to the fact that the graininess function of the time scale affects the error covariance update and the Kalman gain. We thus conclude that the time jumps will affect the estimation performance of the Kalman filter. Moreover, from the error magnitude point of view, it can be asserted that the estimation error increases as the time jump $J$ goes up. Note that the estimation results of the designed Kalman filter differ slightly in multiple sets of simulations due to the presence of random vibration $\omega$. Nevertheless, the estimation effects are generally consistent. For instance, a considerable estimation error always occurs at the time point after the time jump $J$, and the largest estimation error generally arises at $J=6$. The last point worth perceiving is that although the estimation errors at those five points are fairly large, the estimation errors converge so rapidly that they have almost no effect on the estimation of the next state.

\par To further validate the previous conclusions, we design a Kalman filter on the identical time scale $\mathbb{T}_d$ for the reference system (\ref{equ1}). The results are displayed in Fig.~\ref{fig:kf2}. As can be noticed, the five marked points in the right-hand plot are the same as the previous case, locating at the time points after the five jumps, respectively. But unlike before, the estimation errors are within an acceptable range when $t \in\{32,128,212\}$ ($J=2,2,3$ correspondingly) and only become quite alarming when $t \in\{135,224\}$ ($J=4,6$ correspondingly). In fact, it can be inferred from \cite{taousser2016consensus} that for a given dynamical system, there exists an upper bound on the graininess function $\bar\mu$ to ensure that the estimation error is bounded. From simulation verification, we find that the $\bar\mu$ of the OWC system is about $1.4$, while the $\bar\mu$ of the reference system is about $3.1$. This is why the estimation errors in Fig.~\ref{fig:kf1} and Fig.~\ref{fig:kf2} behave differently. Consequently, it can be recognized that the estimation error depends not only on the value of the time jump $J$, but is also closely related to the system itself. Furthermore, the former conclusion that the estimation error is larger at the point with larger $J$ remains valid.

\par According to the numerical example and the above simulation results, the conclusions about the time-scale Kalman filter are described in the following:
\begin{itemize}
    \item Except for some problematic points after time jumps, the state estimation of the Kalman filter on time scales is generally satisfactory.
    \item The estimation error of the problematic point depends on the value of the time jump $J$ and the system itself. 
    \item The estimation error of the problematic point increases as the time jump $J$ increases.
    \item The estimation error of the problematic point converges promptly and has almost no effect on the estimation of the next state.
\end{itemize}
These points demonstrate partially the feasibility of the time-scale Kalman filter. Nevertheless, there are still some limitations of this approach, and more developments are required:
\begin{itemize}
\item Firstly, in our case, the Kalman filter is implemented on a time scale of which the graininess function is bounded, so it is reasonable to obtain bounded estimation errors. Nevertheless, to be rigorous and generalized, sufficient and necessary conditions that enable bounded errors need to be further explored.
\item Secondly, another interesting topic is to find methods that can reduce the estimation errors at the problematic points. For example, a preliminary idea is to ``rescale" the time scale according to the upper bound of the graininess function. 
\item Finally, in this simulation, the time scale is presumed as a priori information. Therefore, it is also a matter of attention to effectively obtain the time scale of the system in a real-world scenario.
\end{itemize}
\section{Conclusion}
\label{conclu}
For the very first time, we apply the time scales theory to a practical application, that is, design a Kalman filter for the vibration-induced intermittent OWC system on time scales. Superior to other traditional methods, the proposed time-scale Kalman filter eases the analysis of a particular system and provides general solutions for systems with non-uniform measurements of arbitrary bounded graininess. Moreover, by revealing the feasibility and current limitations of the time-scale Kalman filter in the OWC application, this study promotes the further development and employment of the time scales theory. For future works, we will investigate the sufficient and necessary conditions for the estimation error to be bounded, and extend our results to generalized intermittent systems under more realistic scenarios.
\bibliographystyle{IEEEtran}
\bibliography{Reference}
\begin{IEEEbiography}[{\includegraphics[width=1in,height=1.25in,clip,keepaspectratio]{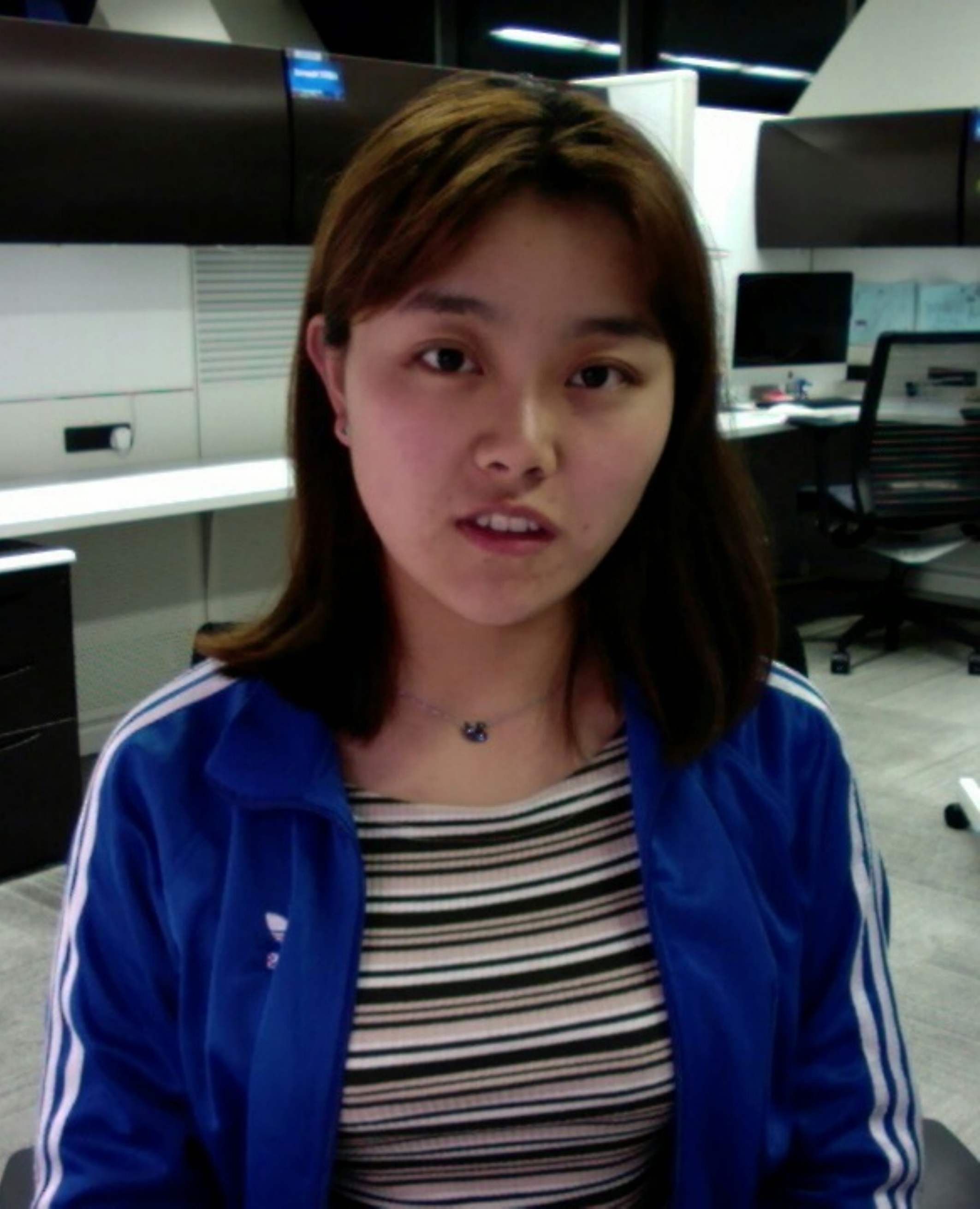}}]{Wenqi Cai}
(Student Member, IEEE) received the B.S. degree in automation engineering from Northeastern University (NEU), China, in 2018; and received the M.S. degree in electrical engineering at King Abdullah University of Science and Technology (KAUST), Saudi Arabia, in 2020, with the research Estimation, Modeling and ANalysis (EMAN) Group and Computer, Electrical and Mathematical Sciences and Engineering (CEMSE) Division. Her research interests include control theory, time scale theory, and reinforcement learning.
\end{IEEEbiography}
\begin{IEEEbiography}[{\includegraphics[width=1in,height=1.25in,clip,keepaspectratio]{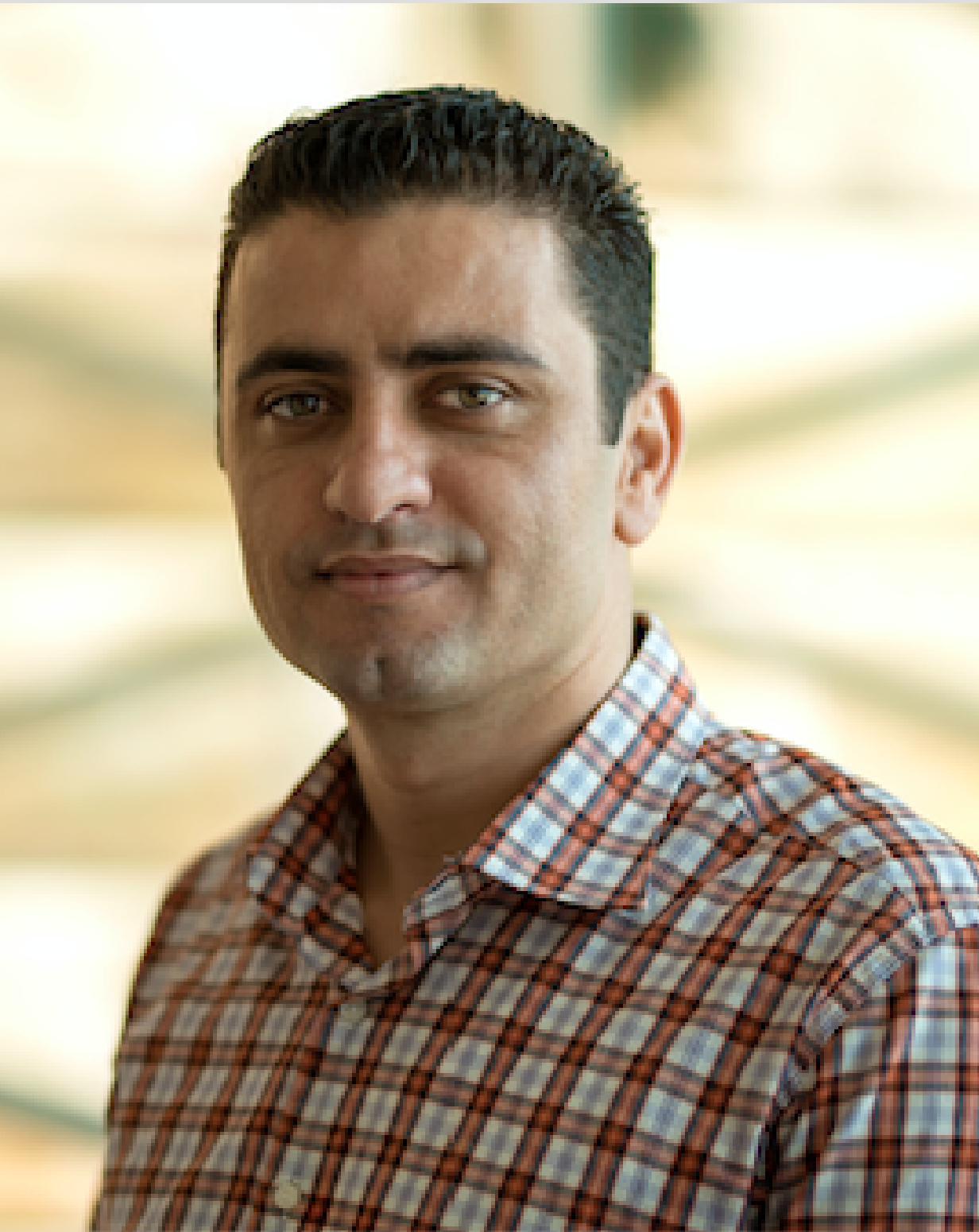}}]{Bacem Ben Nasser}
is currently a Postdoctoral Research Fellow at King Abdullah University of Science and Technology (KAUST), Saudi Arabia, with the research Estimation, Modeling and ANalysis (EMAN) Group. He received his PhD degree in Mathematics from Faculty of Sciences of Sfax, University of Sfax, Tunisia, in 2016. Dr. Bacem joined in 2017, as an Assistant Professor of Applied Mathematics, the University of Kairouan, Higher Institute of Applied Sciences and Technology of Kairouan, Tunisia. 
His current research interests concern the analysis and control design for nonlinear and hybrid dynamical systems including time scales theory and modeling on non-uniform time domains.
\end{IEEEbiography}

\begin{IEEEbiography}[{\includegraphics[width=1in,height=1.25in,clip,keepaspectratio]{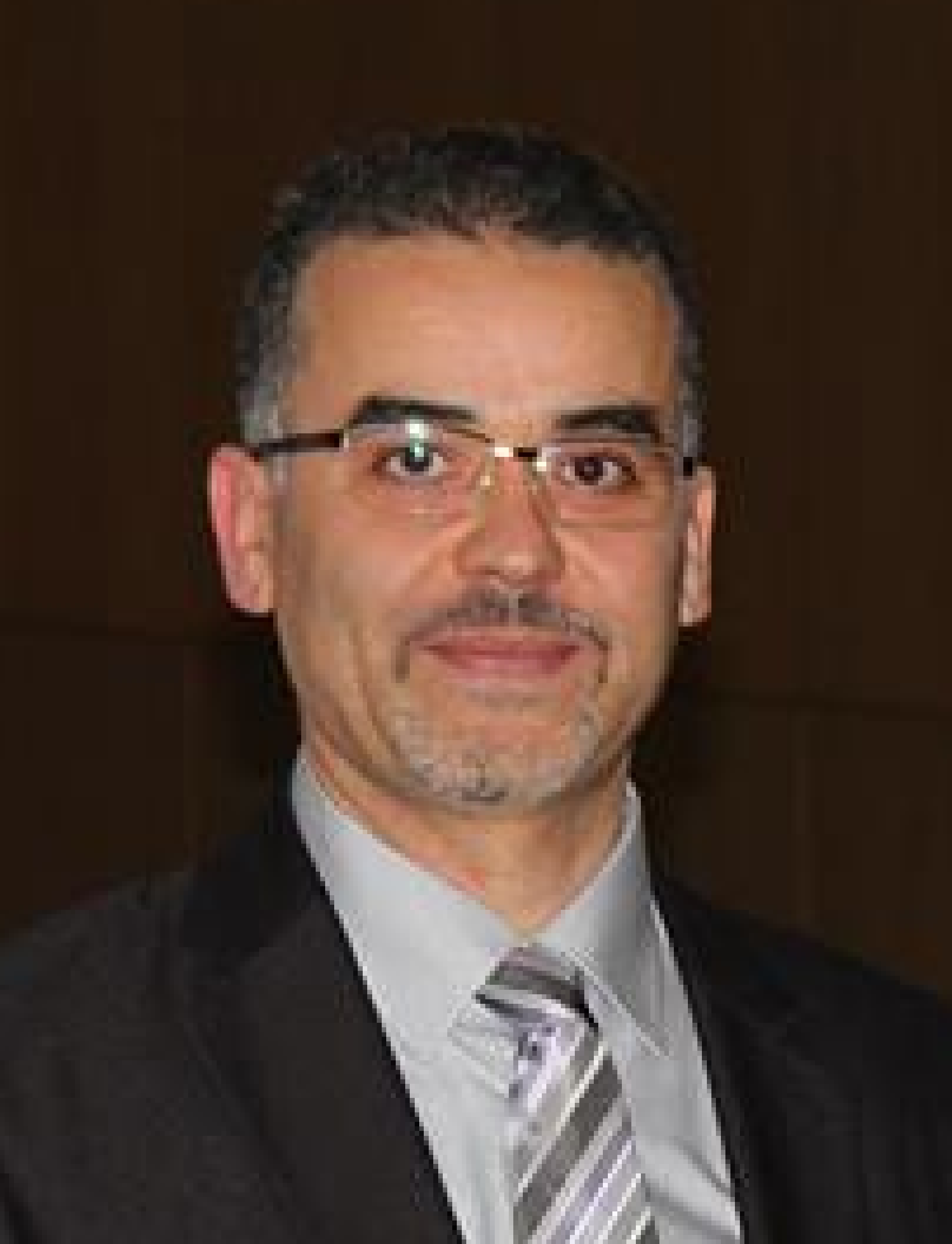}}]{Mohamed Djemaï}
IEEE Senior member, is currently full professor at University Polytechnic Hauts-de France, Valenciennes, France since 2008 and a member of LAMIH Laboratory (CNRS - UMR). Prior to that, Professor Djemaï obtained his B.Sc. in electrical engineering from ENP-Alger, Algeria in 1991, his M.Sc. (DEA) and PhD in Control and Signal System from University of Paris-Sud, France in 1992, and January 1996 respectively. He is member of 2 IFAC TC-2.1 control system, and TC-1.3 on Discrete Event and Hybrid Systems, and 2 IEEE TC on hybrid systems and TC on VSS \&\ SMC. His research interests are mainly related to nonlinear control systems, observation, and fault detection theory including hybrid system, variable structure systems and time scale, with applications to power systems, robotic and vehicles. He published more than 90 journals and 170 Conf. papers in his area of research. He was co-author 03 books. 
\end{IEEEbiography}

\begin{IEEEbiography}[{\includegraphics[width=1in,height=1.25in,clip,keepaspectratio]{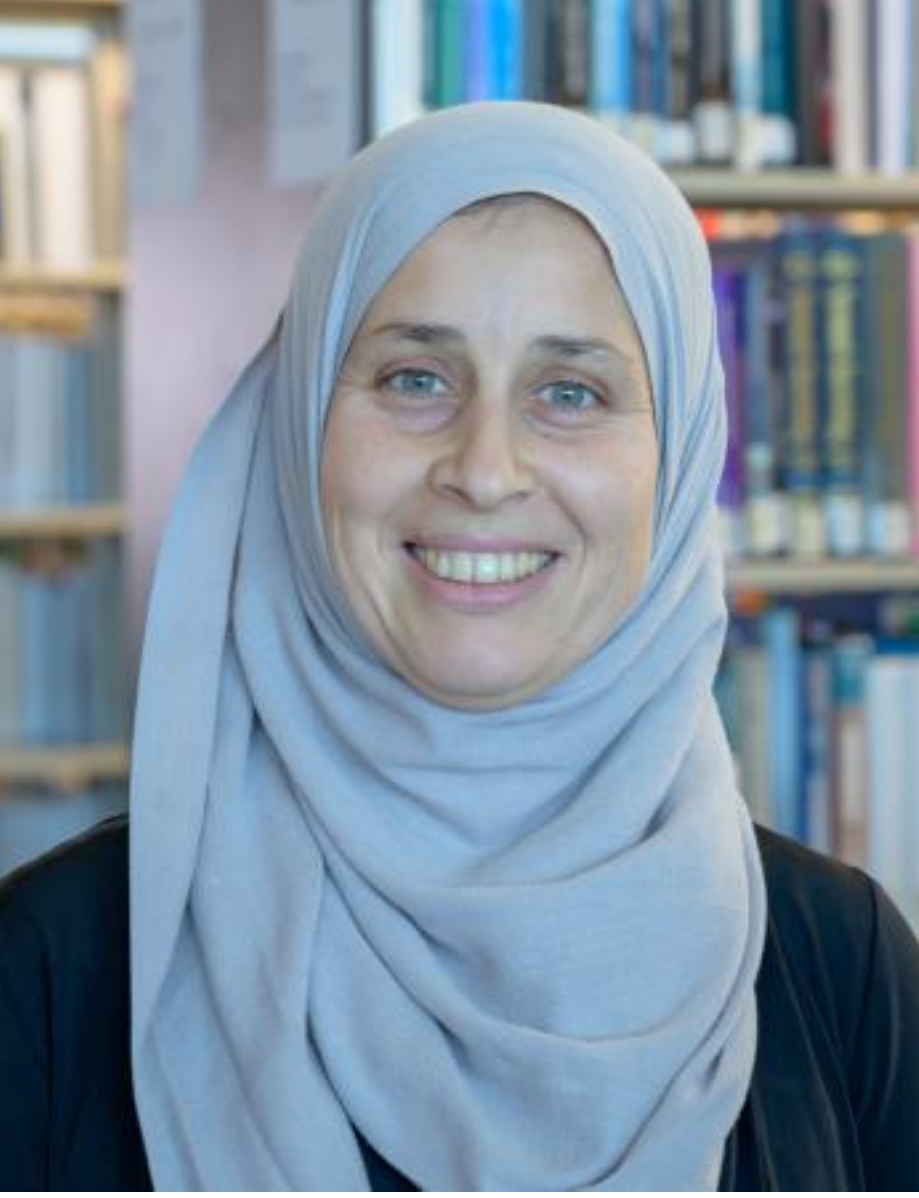}}]{Taous-Meriem Laleg-Kirati} is an associate professor in the division of Computer, Electrical and Mathematical Sciences and Engineering at KAUST and member of the Computational Bioscience Research Center (CBRC). She leads the Estimation Modeling and ANalysis (EMAN) group. She received her Ph.D. in Applied Mathematics in 2008 from INRIA and Versailles University. Professor Laleg’s work is in the general area of mathematical control theory, systems modeling, signal processing and their applications. Her primary research goals are directed towards developing effective estimation methods and algorithms to understand complex systems, extract hidden information and design control and monitoring strategies. She is an IEEE senior member, a member of the IEEE Control Conference Editorial Board, an associate editor of the International Journal of Robust and Nonlinear Control, IEEE Systems Journal and IEEE access Journal and several control conferences including the European Control Conference. She is also member of the international federation of automatic control (IFAC) technical committee on Biological and Medical systems (TC8.2) and Modeling and Control of Environmental Systems (TC8.3).
\end{IEEEbiography}

\end{document}